\PassOptionsToPackage{nopatch}{microtype}
\PassOptionsToPackage{colorlinks=true,citecolor=blue!30!black,linkcolor=black,urlcolor=teal}{hyperref}

\documentclass[
  journal=pasa,
  manuscript=research-paper,
  year=2025,
  volume=YY,
]{cup-journal}

\usepackage{booktabs}
\usepackage{psfrag,color,epsfig}
\usepackage{dcolumn}
\usepackage{graphicx,graphics}	

\usepackage[T1]{fontenc}

\usepackage{amsmath}	
\usepackage{amssymb}	

\usepackage{newtxtext}
\usepackage[nosymbolsc,noamssymbols]{newtxmath}

\usepackage{bm}		
\usepackage{multirow}
\usepackage{tabularx}
\usepackage{tikz}
\usetikzlibrary{matrix,calc}
\usepackage{physics}
\usepackage{xspace}
\usepackage{soul} 

\usepackage{ulem}
\usepackage{xcolor}

\usepackage{hyperref}
\usepackage{orcidlink}
\hypersetup{colorlinks=true,
            citecolor=blue!30!black,
            linkcolor=black,
            urlcolor=teal}

\usepackage{fancyhdr}
\definecolor{midgray}{gray}{0.7}
\newcommand{\Gmidrule}{%
  \arrayrulecolor{midgray}
  \cmidrule[0.6pt]{1-4}
  \arrayrulecolor{black}
}

\def\kk{{\bm{k}}}
\def\qq{{\bm{q}}}

\def\x{\bm{x}}

\def\mpc{{\rm Mpc}}

\def\pp{\hat{\bm{p}}}
\def\emm{{}}

\long\def\/*#1*/{}

\def\kd{\kappa_\delta}
\def\ld{\lambda_\delta}

\title{Probing gravity beyond general relativity with bispectrum multipoles of cosmological tracers: I. Theoretical Foundations}

\author{Sukhdeep Singh Gill \,\orcidlink{0000-0003-1629-3357} }
\affiliation{Department of Physics, Indian Institute of Technology Kharagpur, Kharagpur 721 302, India}
\email[Sukhdeep Singh Gill]{sukhdeepsingh5ab@gmail.com}

\received {dd mm YYYY}
\revised  {dd mm YYYY}
\accepted {dd mm YYYY}
\published{dd mm YYYY}

\keywords{cosmology: theory, cosmology: large-scale structure of the Universe, methods: statistical, galaxies: statistics, gravitation}

\begin{document}
\begin{abstract}
The bispectrum, being sensitive to non-Gaussianity and mode coupling of cosmological fields induced by non-linear gravitational evolution, serves as a powerful probe for detecting deviations from General Relativity (GR). The signatures of modified gravity in the bispectrum are even more pronounced in redshift space, where anisotropies from peculiar velocities provide unbiased information on higher-order properties of gravity. We investigate the potential of all non-zero angular multipoles $B_l^m$ of redshift space bispectrum across all possible triangle configurations to probe degenerate higher-order scalar tensor (DHOST) theory. We show that the higher-order multipoles of the bispectrum with $l=2,4,6$ are more sensitive to the modifications in gravity than the spherically averaged monopole moment $l=0$. These multipoles demonstrate remarkable sensitivity to the higher-order growth history, which varies across triangle configurations, with acute triangles generally being the most sensitive to modification in GR. The values of various multipoles exhibit opposite signs in modified gravity compared to those predicted in GR, which serves as a robust indicator of the deviation from GR. We demonstrate that, unlike $l=2$ and $4$ multipoles, the $l=6$ multipoles with $m\leq 4$ are not affected by the quadratic bias and second-order tidal bias parameters, emphasising the need to leverage their capabilities in analyses. The $(l=6, m > 4)$ multipoles fail to capture the second-order growth, while all $l=8$ multipoles lack any independent information regarding modified gravity in both linear and nonlinear regimes.
\end{abstract}

\section{Introduction}
\label{sec:intro}
The late-time accelerated expansion of the Universe \citep{Riess:1998cb,Perlmutter:1998np} cannot be explained within the framework of General Relativity (GR) coupled with the matter content possessing a positive equation of state. One plausible solution is to introduce an exotic form of fluid with negative pressure, dubbed dark energy, that drives the accelerated expansion \citep{Copeland:2006IJMPD..15.1753C}. The vacuum energy ($\Lambda$) is the simplest candidate of the dark energy. However, there is a huge discrepancy between the theoretically predicted value and the observed value, and its incredibly small magnitude necessitates extreme fine-tuning \citep{Weinberg:1988cp}. Additionally, the standard $\Lambda$CDM ($\Lambda~ +$ cold dark matter) model gives rise to tensions in Hubble constant ($H_0$) and the amplitude of matter fluctuations ($\sigma_8$) between values obtained from direct measurements and those inferred by extrapolating the best-fit parameters from the cosmic microwave background (CMB) data \citep{Efstathiou:2014MNRAS.440.1138E,Abbott:2018PhRvD..98d3526A,Brieden:2023JCAP...04..023B}. A wide variety of dynamic dark energy models have been explored in an effort to address these issues \citep{Sahni:2000IJMPD...9..373S,Peebles:2003RvMP...75..559P,Padmanabhan:2003PhR...380..235P,Copeland:2006IJMPD..15.1753C,Jassal:2005MNRAS.356L..11J,Alam:2004MNRAS.354..275A}. 


An alternative explanation of cosmic acceleration is that the gravity is not well understood on cosmological scales and needs to be modified \citep{Clifton:2012PhR...513....1C}. Any such modification to the theory of gravity must incorporate new degrees of freedom, as the GR is the unique Lorentz invariant low energy theory \citep{Weinberg:1965PhRv..138..988W}. These additional degrees of freedom are associated with new fields, with scalar fields being the simplest candidates. To ensure the Ostrogradsky stability \citep{Ostrogradsky:1850fid,Woodard:2015zca}, the attention is restricted to the class of theories that propagate a single scalar degree of freedom \citep{Langlois:2018dxi,Kobayashi:2019hrl,Kase:2018aps}. The Horndeski theory is the most general scalar-tensor theory that yields equations of motion that are second-order in scalar and metric fields \citep{Horndeski:1974wa}. It also incorporates the Vainshtein screening mechanism \citep{Vainshtein:1972PhLB...39..393V}, which suppresses the force mediated by the additional scalar field on solar system scales \citep{Will:2014LRR....17....4W}. The Horndeski theories were initially considered to be the only viable scalar-tensor formulation that avoids introducing additional ghost degrees of freedom \citep{Deffayet:2011gz,Kobayashi:2011nu}. Beyond the Horndeski framework, however, the theory can be extended to a broader class of scalar-tensor theories, known as De-generate Higher-Order Scalar-Tensor (DHOST) theories \citep{Zumalacarregui:2013pma,Langlois:2015cwa,Gleyzes:2014dya,Gleyzes:2014qga,Crisostomi:2016czh,Achour:2016rkg,BenAchour:2016fzp,Langlois:2018dxi,Kobayashi:2019hrl}. The action of the DHOST theories includes the higher-order terms of second derivatives of a scalar field \citep{Langlois:2015cwa}. The degeneracy in kinetic terms of the scalar field and metric field enables the formulation of viable scalar-tensor theories despite the presence of higher-order equations of motion and avoids the Ostrogradsky instability by maintaining the one scalar propagating degree of freedom \citep{Crisostomi:2016czh}. 


Modifications to gravity alter the growth history of density perturbations and the clustering properties of the large-scale structure (LSS) of the Universe. Measurements of the perturbation growth rate serve as a powerful tool for probing the nature of gravity \citep{Gong:2008PhRvD..78l3010G,Bernardeau:2011JCAP...06..019B,Sharma:2022arXiv220306741S,Brax:2012PhRvD..86f3512B}. These measurements are typically obtained by observing the peculiar velocities of dark matter tracers along the line of sight (LOS) through redshift-space distortions (RSD) \citep{Kaiser:1987qv,Hamilton:1997zq}. A widely adopted approach to test gravity involves the introduction of the gravitational growth index, $\gamma$, which in the framework of the GR is predicted to have a value of approximately $6/11$ \citep{Linder:2007hg,Peebles:1980lssu.book.....P,Fry:1985PhLB..158..211F,Lightman:1990ApJS...74..831L}. Previous studies have not reported any significant deviations of $\gamma$ from the predictions of the GR \citep{Zhao:2019MNRAS.482.3497Z,Gil-Marin:2018MNRAS.477.1604G,Sanchez:2017MNRAS.464.1640S,Grieb:2017MNRAS.467.2085G}. However, to explore the potential signatures of modified gravity more comprehensively, it is necessary to extend these investigations into the regime of non-linear growth, which may provide access to additional information about modified gravity that is not imprinted on $\gamma$ \citep{Yamauchi:2017ibz}. \citet{Yamauchi:2021arXiv210802382Y} have proposed that incorporating non-linear velocity fields alongside non-linear density perturbations provides a robust framework for studying the non-linear features of gravity. They introduced second-order indices that can unveil new insights into modified gravity, capturing non-linear gravitational effects that remain inaccessible to the linear growth index. 

The bispectrum is the lowest-order statistic sensitive to non-Gaussianity induced by non-linear gravitational interaction, making it an ideal tool for investigating the impact of modified gravity on the LSS \citep{Gil-Marin:2017MNRAS.465.1757G,Slepian:2016kfz,Pearson:2018MNRAS.478.4500P,Sugiyama:2018yzo,Sugiyama:2020uil}. It is a function of the closed triangles formed by three $\kk$ modes and captures the coupling between modes. The observed bispectrum of tracers is anisotropic along the LOS direction due to the effects of their peculiar velocities, offering additional sensitivity to deviations from GR. This anisotropy of the redshift space bispectrum can be studied by decomposing it into spherical harmonics $Y_l^m(\hat{p})$  \citep{Scoccimarro:1999ed,Bharadwaj:2020MNRAS.493..594B,Mazumdar:2020MNRAS.498.3975M}. The bispectrum multipole moments $B_l^m$, which are the expansion coefficients of the redshift space bispectrum in $Y_l^m(\hat{p})$ basis, can be directly measured from data. The $B_l^m$ are expected to exhibit distinct signatures due to the altered LSS clustering and mode coupling introduced by the modified gravitational evolution. 

The primary objective of this study is to examine how modifications in gravitational theory, at both linear and non-linear scales, influence the bispectrum multipole signal. While most of the earlier studies are either restricted to the spherically averaged bispectrum ($B_0^0$) only \citep{Hirano:2018uar,Crisostomi:2019vhj,Borisov:2009PhRvD..79j3506B,Bernardeau:2011JCAP...06..019B,Bartolo:2013ws,Takushima:2013foa,Bellini:2015wfa,Burrage:2019afs,Lewandowski:2019txi,Borisov:2009PhRvD..79j3506B,Gil-Marin:2011JCAP...11..019G,Hellwing:2013rxa,Bose:2018zpk,Bose:2020JCAP...02..025B,Yamauchi:2017ibz,Yamauchi:2021arXiv210802382Y,Dinda:2018eyt,Munshi:2019csw,Namikawa:2018erh,Namikawa:2018bju} or a few triangle configurations \citep{Bose:2020JCAP...02..025B,Sugiyama:2023MNRAS.524.1651S}, \citet{Sugiyama:2023MNRAS.523.3133S} has incorporated $l=2$ multipoles also in constraining the modified gravity. Here, for the first time, we consider all possible non-zero multipoles $B_l^m$  across all possible unique triangle configurations to systematically explore their sensitivity to deviation from GR, with a particular focus on the DHOST theories. We investigate the distinctive signatures in $B_l^m$ that indicate departures from GR and identify the triangle configurations most sensitive to the effects of modified gravity. The influence of tracer bias on $B_l^m$, which may entangle with the modified gravity signals, is also analysed. Specifically, we examine the impact of linear and quadratic bias parameters on the $B_l^m$ and determine the configurations most affected. This comprehensive analysis seeks to identify robust observational signatures of modified gravity in $B_l^m$, complementing existing cosmological probes and offering new insights into the fundamental nature of gravity.

A brief outline of the paper is as follows. Section \ref{sec:delta_in_mg} presents the evolution of density and velocity fluctuations in the DHOST theory, including the modelling of perturbation growth. Section \ref{sec:rsd_bs} presents the redshift space bispectrum of biased tracers and their multipole expansion. Section \ref{sec:results} presents the results, and Section \ref{sec:summary} presents the Summary and Discussion. Results for some of the multipole moments, which are not included in the main text, have been presented in the \ref{sec:appendix}.






\section{Density fluctuations in Modified gravity theories}
\label{sec:delta_in_mg}
\subsection{Matter Field}

The dark matter behaves as a pressure-less perfect fluid with no vorticity on large scales \citep{Bernardeau:2001qr}. The fluctuations in its density $\delta(\x,\eta)$ and velocity field ${\bm v}(\x,\eta)$ are described by fluid equations, 
\begin{align}
\label{eq:fluid1}
\dot{\delta} + a^{-1} \nabla. \left[ (1 + \delta) {\bm v} \right] &= 0\,~ , \\
\dot{\theta} + H \theta + a^{-1} \nabla. \left[ ({\bm v} . \nabla) {\bm v} \right] &= -a^{-1} \nabla^2 \phi\,~ ,
\label{eq:fluid2}
\end{align}
where $a$, $H=\dot{a}/a=\frac{1}{a}\frac{da}{d\eta}$, and $\theta=\nabla.\bm{v}$ are the scale factor, Hubble parameter, and velocity divergence, respectively. The potential $\phi(\x,\eta)$ accounts for the influence of gravitational interactions under the assumption that matter and gravity are minimally coupled.  We consider the DHOST theories \citep{Langlois:2017mxy,Langlois:2017mdk}, which is a broad class of modified gravity models.  The gravitational potential satisfies the modified Poisson equation in the DHOST theories under the quasi-static approximation (e.g., \citep{Pace:2020qpj,Hirano:2018uar,Hirano:2020dom} ), 


\begin{equation}
\label{eq:potential_ev}
\frac{\nabla^2 \phi}{a^2 H^2} = \nu_0 \delta + \nu_1 \frac{\dot{\delta}}{H} + \nu_2 \frac{\ddot{\delta}}{H^2} + \frac{\nabla^2 S_\phi^{\text{NL}}}{a^2 H^2},
\end{equation}
where $\nu_0,\nu_1,\nu_2$ are time-dependent functions, and their relation to dark energy effective field theory (EFT) parameters is given in \citep{Yamauchi:2021arXiv210802382Y}. The source term $S_\phi^{\text{NL}}(\x,\eta)$  arises from the equation of motion of the scalar field and encapsulates the contributions of non-linear interactions only to the evolution of potential perturbations. The set of these equations is solved perturbatively by expanding $\delta, \theta$, and $\phi$ fields in terms of successive orders in $\delta_1$ (i.e. $\{\delta,\theta,\phi\}\propto \sum_n\mathcal{O}(\delta_1^n)$), where $\delta_1$ represents the linear order fluctuations in the density \citep{Goroff:1986ApJ...311....6G}. The perturbative expansion of the source term $S_\phi^{\text{NL}}$ in Equation~\ref{eq:potential_ev} is required to solve this system of equations. Its leading contribution arises at the second order  $\mathcal{O}(\delta_1^2)$ and is given by \citep{Sugiyama:2023MNRAS.523.3133S},
\begin{equation}
    \dfrac{\nabla^2 S_{\phi_2}^{\text{NL}}}{a^2H^2} = \tau_1W_1(\x) - \tau_2W_2(\x) ~,
\end{equation}
where $\tau_1,\tau_2$ are time-dependent functions related to the EFT parameters. The spatially dependent functions $W_1$ and $W_2$ are expressed in terms of linear density perturbations $\delta_1$ as,
\begin{equation}
\begin{aligned}
W_1(\x) &= \left[\delta_1(\x)\right]^2 
+ \left[\frac{\partial_i}{\partial^2} \delta_1(\x)\right] 
\left[\partial_i \delta_1(\x)\right], \\
W_2(\x) &= \left[\delta_1(\x)\right]^2 
- \left[\frac{\partial_i \partial_j}{\partial^2} \delta_1(\x)\right]^2.
\end{aligned}
\end{equation}

The  time evolution of the density perturbations, obtained by combining Equations~(\ref{eq:fluid1}),(\ref{eq:fluid2}) and eliminating $\Phi$ using Equation~(\ref{eq:potential_ev}), in the Fourier space is,
\begin{equation}\label{eq:density_ev}
    \ddot{\Delta} + \left(2 + \varsigma\right) H \dot{\Delta} - \frac{3}{2} \Omega_m \Xi H^2 \Delta = H^2 S_\Delta^{\text{NL}},
\end{equation}
where, ${\Delta}(\kk,\eta)$ is the Fourier transform of $\delta(\x,\eta)$, and parameters $\varsigma=(2\nu_2-\nu_1)/(1-\nu_2)$ and $\Xi = 2\nu_0/[3\Omega_m(1-\nu_2)]$ are functions of time. This equation provides the evolution of density fluctuations for a particular DHOST gravity model specified by the set of parameters \{$\nu_0,\nu_1,\nu_2,\tau_1,\tau_2$\} \citep{Sugiyama:2023MNRAS.523.3133S}. In the linear regime, the source term $S_\Delta^{\text{NL}}$ vanishes, and we get the first order growing mode solution $\Delta_1(\kk,\eta)= D(\eta)\Delta_L(\kk)$, where the time-dependent function $D(\eta)$ is the linear growth factor, and $\Delta_L(\kk)$ is the initial density fluctuations. It indicates that the perturbations grow independently for all $\kk$ modes. The evolution equation at first order can be reinterpreted in terms of linear growth rate, $f=\frac{d\ln D}{d\ln a}$, as,
\begin{equation}\label{eq:f_ev}
    \dfrac{df}{d\ln a}+\left(2+\varsigma+\dfrac{d\ln H}{d\ln a}\right)f+f^2-\dfrac{3}{2}\Omega_m\Xi=0~.
\end{equation}

The linear perturbations in the velocity field, calculated using the continuity equation Equation~(\ref{eq:fluid1}), is $\Theta_1(\kk,\eta)=-aHfD\Delta_L(\kk,\eta)$, where $\Theta(\kk,\eta)$ is the Fourier conjugate of the velocity divergence $\theta(\x,\eta)$. We proceed further to compute the second-order solutions, where the non-linear source term $S_\Delta^{\text{NL}}$ (in Equation~\ref{eq:density_ev}) also contributes, as expressed by 
\begin{equation}\label{eq:source_2nd}
\begin{aligned}
    S_{\Delta_2}^{\text{NL}}(\kk,\eta) &= D^2\int \dfrac{d^3q_1}{(2\pi)^3}\int \dfrac{d^3q_2}{(2\pi)^3} (2\pi)^3\delta_D(\kk-\qq) ~\\& \times[\sigma_1 C_{12}^\kappa+\sigma_2 C_{12}^\lambda]~\Delta_L(\qq_1)~\Delta_L(\qq_2)~,
\end{aligned}
\end{equation}
where, $\qq=\qq_1+\qq_2$, $\delta_D$ is the Dirac delta function, $\sigma_1$ and $\sigma_2$ are time-dependent functions \citep{Hirano:2020dom,Sugiyama:2023MNRAS.523.3133S},
\begin{equation}
\begin{aligned}
\sigma_1 &= (1-\nu_2)^{-1}\left( 2f^2+\dfrac{3}{2}\Omega_m\Xi-\varsigma f+\tau_1 \right) , \\
\sigma_2 &= (1-\nu_2)^{-1} \left( -f^2+\tau_2 \right) ,
\end{aligned}
\end{equation}
and the coupling variables $C^{\kappa,\lambda}_{ij}$,
\begin{equation}
\begin{aligned}
C^{\kappa}_{ij} =&~ 1+ (\hat{q}_i.\hat{q}_j) \dfrac{q_i^2+q_j^2}{2q_iq_j}, \\
C^{\lambda}_{ij} =&~ 1- (\hat{q}_i.\hat{q}_j)^2.
\end{aligned}
\end{equation}

The second-order perturbations in density (using Equations~\ref{eq:density_ev} and \ref{eq:source_2nd}) and velocity field (using Equation~\ref{eq:fluid1}) are given, respectively, by,
\begin{equation}\label{eq:2nd_order_density}
\begin{aligned}
\Delta_2(\kk,\eta) =& ~D^2 \int \dfrac{d^3q_1}{(2\pi)^3}\int \dfrac{d^3q_2}{(2\pi)^3} (2\pi)^3\delta_D(\kk-\qq)\\\times &~F_2^{\emm}(\qq_1,\qq_2,\eta)~\Delta_L(\qq_1)~\Delta_L(\qq_2)~,
\end{aligned}
\end{equation}
\begin{equation}
\begin{aligned}
\Theta_2(\kk,\eta) =& -aHfD^2 \int \dfrac{d^3q_1}{(2\pi)^3}\int \dfrac{d^3q_2}{(2\pi)^3} (2\pi)^3\delta_D(\kk-\qq)\\\times &~G_2^{\emm}(\qq_1,\qq_2,\eta)~\Delta_L(\qq_1)~\Delta_L(\qq_2)~.
\end{aligned}\label{eq:2nd_order_velocity}
\end{equation}



The second-order kernels $F_2^{\emm}$ and $G_2^{\emm}$ are \citep{Hirano:2020dom,Takushima:2013foa,Takushima:2015iha,Crisostomi:2019vhj,Lewandowski:2019txi},
\begin{equation}
\begin{aligned}\label{eq:2nd_order_matter_kernels}
F_2^{\emm}(\qq_i,\qq_j,\eta) =& ~\kappa_{\delta} C^{\kappa}_{ij}-\dfrac{2}{7}\lambda_\delta C^{\lambda}_{ij} \\
G_2^{\emm}(\qq_i,\qq_j,\eta) =& ~\kappa_\theta C^{\kappa}_{ij}-\dfrac{4}{7}\lambda_\theta C^{\lambda}_{ij}
\end{aligned}
\end{equation}
The evolution of the density field at the second-order can be reinterpreted in the form of the evolution of non-linear growth rates $\kappa_\delta$ and $\lambda_\delta$, using Equations~(\ref{eq:2nd_order_density}) and (\ref{eq:source_2nd}) in Equation~(\ref{eq:density_ev}),
\begin{equation}
\begin{aligned}\label{eq:kd_ev}
&\dfrac{d^2{\kappa}_\delta}{d(\ln a)^2}+\left(4f+2+\varsigma+\dfrac{d\ln H}{d\ln a}\right)\dfrac{d{\kappa}_\delta}{d\ln a} + \left(2f^2+\dfrac{3}{2}\Omega_m\Xi\right)\kd \\&\hspace{.33\textwidth}= \sigma_1 ~,
\end{aligned}
\end{equation}
\begin{equation}
\begin{aligned}\label{eq:ld_ev}
&\dfrac{d^2{\lambda}_\delta}{d(\ln a)^2}+\left(4f+2+\varsigma+\dfrac{d\ln H}{d\ln a}\right)\dfrac{d{\lambda}_\delta}{d\ln a} + \left(2f^2+\dfrac{3}{2}\Omega_m\Xi\right)\ld  \\&\hspace{.33\textwidth}=\dfrac{7}{2} \sigma_2~.
\end{aligned}
\end{equation}

The non-linear growth rates of the velocity divergence field $\kappa_\theta$ and $\lambda_\theta$ can be determined from $\kd$ and $\ld$, and the relations obtained using Equations~(\ref{eq:2nd_order_density}) and $(\ref{eq:2nd_order_velocity})$ in the Equation~(\ref{eq:fluid1}) are \citep{Hirano:2020dom,Sugiyama:2023MNRAS.523.3133S},

\begin{equation}
\begin{aligned}\label{eq:k_th}
\kappa_\theta =&~ 2\kappa_\delta\left[ 1+\dfrac{1}{2f}  \dfrac{d\ln\kappa_\delta}{d\ln a}\right]-1~, \\
\lambda_\theta =&~ \lambda_\delta\left[ 1+\dfrac{1}{2f}  \dfrac{d\ln\lambda_\delta}{d\ln a}\right].
\end{aligned}
\end{equation}

\subsection{Modelling Growth of Perturbations in DHOST Theories}\label{sec:model}

The Equations~(\ref{eq:f_ev}), (\ref{eq:kd_ev}), and (\ref{eq:ld_ev}) provide a complete description of first and second-order evolution of density perturbations within the framework of the DHOST gravity theories. The specific model of gravity is characterised by the parameters $\varsigma$, $\Xi$, $\sigma_1$, and $\sigma_2$. These parameters are, in turn, determined by the EFT parameters \{$\nu_0$, $\nu_1$, $\nu_2$, $\tau_1$, $\tau_2$\}, which encapsulate the modifications introduced by the DHOST model. In the standard $\Lambda$CDM model with GR, the parameters take values \{$\nu_0=\frac{3}{2}\Omega_m$, $\nu_1=0$, $\nu_2=0$, $\tau_1=0$, $\tau_2=0$\}, hence $\varsigma=0$, $\Xi=1$, $\sigma_1=2f^2+\frac{3}{2}\Omega_m$, and $\sigma_2=-f^2$ and the Equations~(\ref{eq:f_ev}), (\ref{eq:kd_ev}) and (\ref{eq:ld_ev}) lead to $f=\Omega_m^{6/11}$, $\kd=1$ and $\ld=\Omega_m^{3/572}$ \citep{Bouchet:1995A&A...296..575B,Bernardeau:2001qr,Yamauchi:2017ibz}. The second-order growth rates of the velocity divergence field obtained using Equation~(\ref{eq:k_th}) are $\kappa_\theta=1$ and $\lambda_\theta=\Omega_m^{15/1144}$. In the Horndeski gravity, the constraints are limited to \{ $\nu_1=0$, $\nu_2=0$, $\tau_1=0$\} only, and the remaining parameters \{$\nu_0, \tau_2$\} are free to take the values, allowing deviation from the standard model of gravity. This results in  $\varsigma=0$, $\sigma_1=2f^2+\frac{3}{2}\Omega_m$ and $\{\Xi,\sigma_2\}$ depart from their standard $\Lambda$CDM values \citep{Yamauchi:2017ibz,Takushima:2013foa,Sugiyama:2023MNRAS.524.1651S}. Consequently, $\kd=\kappa_\theta=1$, while $\ld$ (and $\lambda_\theta$) captures the modification to gravity. In the general case of DHOST theories, all the growth rates $f,\kd,\ld$ (also $\kappa_\theta,\lambda_\theta$) diverge from those predicted by standard $\Lambda$CDM+GR \citep{Crisostomi:2019vhj,Lewandowski:2019txi,Hirano:2018uar}. It suffices to model these growth rates to study the perturbations in the DHOST framework.

We adopt the widely used model of linear growth rate as \citep{Wang:1998ApJ...508..483W},
\begin{equation}\label{eq:model_1st}
    \begin{aligned}
        f = \Omega_m^\gamma(a) ~ , 
    \end{aligned}
\end{equation}
where $\gamma$ is the linear growth index and it takes value $\gamma_0=6/11$ for the standard $\Lambda$CDM + GR theory. It has been utilised in studying the growth of linear perturbations. Recent works by \citet{Yamauchi:2017ibz} and \citet{Yamauchi:2021arXiv210802382Y} have demonstrated that the second-order growth rates in the DHOST theories can similarly be expressed as a power of $\Omega_m$. Following their formulation, we proceed with the model,
\begin{equation}\label{eq:model_2nd}
    \begin{aligned}
        \kappa_\delta = \Omega_m^\psi(a)~ , \hspace{.3cm} \lambda_\delta= \Omega_m^\xi(a) .
    \end{aligned}
\end{equation}
where $\psi$ and $\xi$ are both the second-order growth indices and take values 0 and $3/572$, respectively, in the $\Lambda$CDM + GR case.  Note that we take the background evolution of $\Omega(a)$ assuming the standard GR, given by,
\begin{equation}
    \begin{aligned}
        \Omega_m(a) = \dfrac{\Omega_{m_0} a^{-3}}{\Omega_{m_0} a^{-3}+1-\Omega_{m_0}} .
    \end{aligned}
\end{equation}
In principle, the time evolution of $\Omega_m(a)$ in the DHOST theories differs from that in GR. Nevertheless, this deviation is suppressed by the factor $(1-\Omega_{\rm GR})$ \citep{Yamauchi:2021arXiv210802382Y}.

\subsection{Biased Tracers in Redshift Space}

The observables, such as galaxy distribution, Lyman-$\alpha$, 21-cm signal, etc, trace the dark matter with some bias. The density perturbations of these biased tracers $\delta^{(b)}(\x,\eta)$ are modelled using a Taylor series expansion of matter fluctuations $\delta(\x,\eta)$, up to second order as follows \citep{Desjacques:2016bnm},
\begin{equation}
    \delta^{(b)} = b_1\delta + \dfrac{b_2}{2}\delta^2 + b_{s^2}s_{ij}^2~,
\end{equation}
where $b_1$ and $b_2$ are the local linear and quadratic bias parameters, respectively, and $b_{s^2}$ is the second-order
tidal bias. The source of tidal perturbations $s_{ij}^2$ is expressed as,
\begin{equation}\label{eq:tidal_force}
    s_{ij}^2 = \left[ \dfrac{\partial_i\partial_j}{\partial^2} \delta \right]^2 - \dfrac{1}{3} \delta^2~.
\end{equation}
The tracers are observed in redshift space. The observed redshift includes the contribution of peculiar velocity also in addition to the Hubble expansion, and the estimated distance $\tilde{\bm{x}}$ is distorted along the LOS direction $\hat{n}$, expressed as $\tilde{\bm{x}} =  \x+\frac{\bm{v}.~\hat{n}}{aH}\hat{n}$ \citep{Kaiser:1987qv}. The $n^{\rm th}$ order perturbations in the tracer density field in redshift space are given by \citep{Scoccimarro:1999ed},
\begin{equation}\label{eq:delta_bs}
\begin{aligned}
\tilde{\Delta}^{(b)}_n(\kk,\eta) =&~ D^n \int \dfrac{d^3q_1}{(2\pi)^3}\dots\int \dfrac{d^3q_n}{(2\pi)^3} (2\pi)^3\delta_D(\kk-\qq)\\\times &~Z_n(\qq_1,\dots,\qq_n,\eta)~\Delta_L(\qq_1)\dots\Delta_L(\qq_n).
\end{aligned}
\end{equation}

The first and second-order redshift space kernels, assuming the peculiar velocity of the tracers follow the matter velocity $\bm{v}^{(b)}=\bm{v}$ are,
\begin{equation}\label{eq:Z1_Z2}
\begin{aligned}
Z_1(\qq,\eta) &= ~b_1 + f~(\hat{q}.\hat{n})^2~,\\ ~
Z_2(\qq_1,\qq_2,\eta)&= ~b_1F_2^{\emm}(\qq_1,\qq_2,\eta)+\dfrac{1}{2}b_2 + b_{s^2}\bigg[(\hat{q}_1.\hat{q}_2)^2\\& - \dfrac{1}{3}\bigg]+f(\hat{q}.\hat{n})^2 G_2(\qq_1,\qq_2,\eta) + \dfrac{f(\qq.\hat{n})}{2}\\&\times\left[  \dfrac{(\hat{q}_1.\hat{n})}{q_1}Z_1(\qq_2,\eta)  +  \dfrac{\hat{q_2}.\hat{n}}{q_2} Z_1(\qq_1,\eta)\right].
\end{aligned}
\end{equation}

\section{Redshift Space Bispectrum of Biased Tracers}
\label{sec:rsd_bs}
The bispectrum is defined as,
\begin{equation}
    B(\kk_1,\kk_2,\kk_3)=\langle~ \tilde{\Delta}^{(b)}(\kk_1) \tilde{\Delta}^{(b)}(\kk_2) \tilde{\Delta}^{(b)}(\kk_3)~\rangle
\end{equation}
with the condition that $\kk$ modes form a closed triangle, $\kk_1+\kk_2+\kk_3=0$. The leading order contribution is given by (using Equation~\ref{eq:delta_bs}) \citep{Scoccimarro:1999ed,Mazumdar:2020MNRAS.498.3975M},
\begin{equation}\label{eq:2ptbs}
    \begin{aligned}
B(\kk_1,\kk_2,\kk_3)&=2D^4Z_2(\kk_1,\kk_2,\eta)Z_1(\kk_1,\eta)Z_1(\kk_2,\eta)\\ & \times P_L(k_1)P_L(k_2) + \text{cyc.}~,
\end{aligned}
\end{equation}
where $P_L(k)=\langle |\Delta_L(k)|^2\rangle$, is the linear matter power spectrum. We have used the Boltzmann code \texttt{CLASS} \citep{Blas:2011rf} to compute the linear power spectrum $P_L(k)$. The redshift space bispectrum is a function of five parameters that define the triangle's shape, size, and orientation with respect to the LOS. We opt for the parametrisation prescribed by \citet{Bharadwaj:2020MNRAS.493..594B}. For a triangle with $k_1\geq k_2\geq k_3$, its size is parameterised by the largest side $k_1$, shape by two dimensionless parameters $(\mu,t)$,
\begin{equation}
\mu =- \frac{\kk_1 \cdot \kk_2}{k_1 k_2}, \hspace{0.5cm} \, t = \frac{k_2}{k_1} \,,
\label{eq:shape}
\end{equation}
and orientation with respect to the LOS using a unit vector $\hat{p}$.  Assuming the fixed LOS along $\hat{z}$, the cosines of angles between the LOS and triangle sides are related to $\hat{p}$ by, 
\begin{eqnarray}
&\mu_{1} & = p_z \,,  \hspace{0.5 cm} \, 
\mu_{2} = -\mu p_z + \sqrt{1-\mu^2}p_x~, \nonumber \\
&\mu_{3}& = \dfrac{-[(1-t\mu)p_z+t\sqrt{1-\mu^2}p_x]}{\sqrt{1-2t\mu+t^2}}~,
\label{eq:orient}
\end{eqnarray}
where, $\mu_i=\hat{k}_i.\hat{z}$. We use the Equation~(\ref{eq:Z1_Z2}) in (\ref{eq:2ptbs}) and rewrite the expression for bispectrum as \citep{Mazumdar:2020MNRAS.498.3975M},
\begin{equation}\label{eq:final_bs}
    \begin{aligned}
        B(k_1,\mu,t,\hat{p}) &= 2D^4\bigg\{ RG^{\emm}_{12}+S_{12}\Big[b_1\Delta G^{\emm}_{12}+\dfrac{b_2}{2} + \dfrac{b_{s^2}}{3}\\&\times\Big(2-3C^\lambda_{12}\Big)\Big] -T_{12} \bigg\}  P_L(k_1)P_L(k_2) +\text{cyc.}
    \end{aligned}
\end{equation}
Here, the term $(2-3C^\lambda_{12})/3$ represents the scale-dependent tidal force (Equation~\ref{eq:tidal_force}). We have used the compact notation for the kernels, $G_{ij}^{\emm}=G_2^{\emm}(\qq_i,\qq_j,\eta)$, etc. and defined the function $\Delta G^{\emm}_{ij}$ as, $\Delta G^{\emm}_{ij} = F_{ij}^{\emm} - G_{ij}^{\emm}$. For a specific model of gravity, the kernels depend on the configuration of triangle shapes via coupling parameters, $C^{\kappa,\lambda}_{ij}$. These are expressed in terms of shape parameters $(\mu,t)$ as follows,
\begin{equation}
    \begin{aligned}
        C^\kappa_{12} &=1-\dfrac{\mu(1+t^2)}{2t}, \hspace{.25cm} C^\kappa_{23} = 1+\dfrac{\mu-t}{2t}\left(\dfrac{t^2}{1+t^2-2\mu t}+1\right)\\
        C^\kappa_{31} &= 1+\dfrac{(\mu t-1)^2}{2}\left( \dfrac{1}{1+t^2-2\mu t}+1\right)\\
               C^\lambda_{12} &=1-\mu^2, \hspace{.15cm} C^\lambda_{23} = \dfrac{1-\mu^2}{1+t^2-2\mu t},  \hspace{.15cm}
        C^\lambda_{31} = \dfrac{t^2(1-\mu^2)}{1+t^2-2\mu t}
    \end{aligned}
\end{equation}

The orientation ($\hat{p}$) dependence of the redshift space bispectrum is fully quantified by the  functions $R(\hat{p}, b_1,f)$, $S_{ij}(\hat{p}, b_1,f)$ and $T_{ij}(\hat{p}, b_1,f, \mu, t)$ given by,
\begin{align}
    R &= (b_1 + f \mu_1^2)(b_1 + f \mu_2^2)(b_1 + f \mu_3^2)~,\\
    S_{ij} &= (b_1 + f \mu_i^2)(b_1 + f \mu_j^2)~,\\
    T_{ij} &= ~ -\frac{f}{2} (b_1 + f\mu_i^2)(b_1 + f \mu_j^2)(1+t^2-2\mu t)^{1/2} \nonumber \\& \times (\mu_i+\mu_j) \left[ \mu_i (b_1 + f \mu_j^2) + \mu_j t^{-1} (b_1 + f \mu_i^2) \right].
\end{align}

These functions encapsulate the linear growth factor $f$ and linear bias $b_1$ only. Sensitivity of the bispectrum to non-linear growth rates $(\kd,\ld)$ enters through the kernels $G^{\emm}_{ij}$ and $\Delta G^{\emm}_{ij}$ (Equation~\ref{eq:2nd_order_matter_kernels}). To study the anisotropy of the bispectrum, we decompose it into spherical harmonic $Y_l^m(\hat{{p}})$ basis and express the bispectrum multipole moments ${B}_l^m(k_1,\mu,t )$ as \citep{Scoccimarro:1999ed,Bharadwaj:2020MNRAS.493..594B,Mazumdar:2020MNRAS.498.3975M},
\begin{equation}
    {B}_l^m(k_1,\mu,t )=\sqrt{\dfrac{2l+1}{4\pi}}{\int[Y_l^m(\hat{{p}})]^*~B(k_1,\mu,t,\hat{p})~d\Omega_{\hat{{p}}}},
\label{eq:bs_lm}
\end{equation}
The multipole expansion of the functions $R$, $S_{ij}$, and $T_{ij}$, which arises by putting Equation~(\ref{eq:final_bs}) in (\ref{eq:bs_lm}), is defined in a similar manner. The expressions for their multipoles are given in \citep{Bharadwaj:2020MNRAS.493..594B} (see Equations 26-29, 31-33, A1-A7 therein) and \citep{Mazumdar:2020MNRAS.498.3975M} (see Equations A1-A22, B1-B65 therein). Note that the definitions of $R$, $S_{ij}$, and $T_{ij}$ adopted in this work differ slightly from those in their studies. Therefore, to ensure consistency, the expressions for $R$, $S_{ij}$, and $T_{ij}$ in \citep{Bharadwaj:2020MNRAS.493..594B} and \citep{Mazumdar:2020MNRAS.498.3975M} need to be multiplied by $b_1^3$, $b_1^2$, and $b_1^4$, respectively. 

We compute the ratio of all non-zero bispectrum multipoles (Equation~\ref{eq:bs_lm}) for varying values of $\gamma,\psi,\xi$ relative to those obtained under the assumption of standard GR ($\gamma_{\rm GR}=6/11, \psi_{\rm GR}=0, \xi_{\rm GR}=3/572$), considering all possible triangle configurations. We restrict the values of $\gamma,\psi,\xi$ that satisfy the constraints given in \citep{Sugiyama:2023MNRAS.523.3133S}. They have placed bounds on the modified growth parameters $\xi_f=\log_{\Omega_m}(f/\kd)$, $\xi_s=\log_{\Omega_m}(\kappa_\theta/\kd)$ and $\xi_t=\log_{\Omega_m}(\lambda_\theta/\kd)$ using data from the BOSS survey, finding that $-0.907<\xi_f<2.447$, $-0.504<\xi_s$ and $-1.655<\xi_t$ with $95\%$ confidence. The results are discussed in the next Sections.

\section{Results}
\label{sec:results}

The bispectrum $B(k_1,\mu,t,\hat{p})$ (Equation~\ref{eq:final_bs}) exhibits an approximate power-law dependence on $k_1$ within the scales of interest. Consequently, the key features of the bispectrum multipole ratio between different gravitational theories remain consistent across $k_1$. Thus, showing the results for a single value of $k_1$ is sufficient to capture the essential behaviour. Here, we present the results for $k_1=0.1$ Mpc$^{-1}$ and adopt redshift $z=0.61$. Unless stated otherwise, we use fiducial values for the bias parameters $b_1=1.2$, $b_2=0$, and $b_{s^2}=0$ \citep{Nandi:2024NewA..11302292N}. The following subsections present a detailed examination of the sensitivity of bispectrum multipoles to the linear and non-linear growth indices.

\subsection{Sensitivity to linear growth}
\begin{figure*}
    \centering 
    \includegraphics[width=1\textwidth]{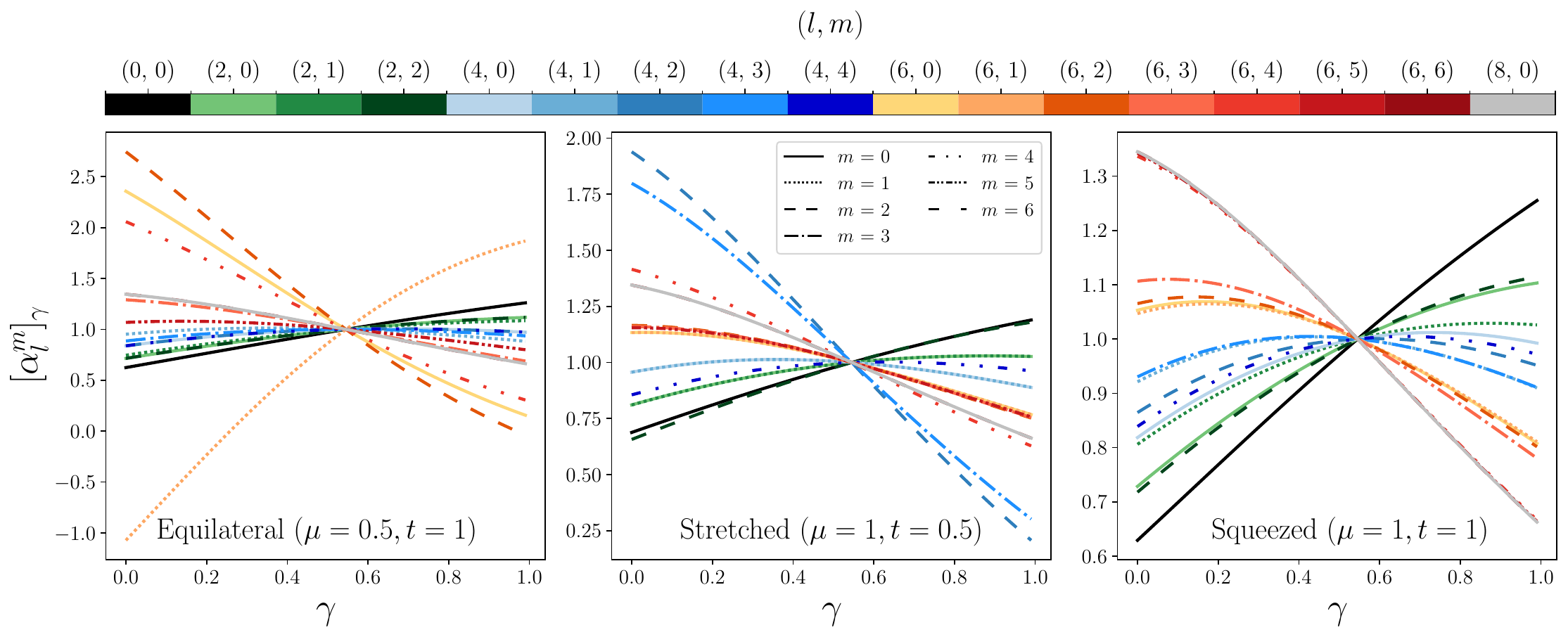} 
    \caption{Sensitivity of $B_l^m$ to the linear growth index $\gamma$. It shows the variation of the ratio $[\alpha_l^m]_\gamma=[B_l^m]_\gamma/[B_l^m]_{\rm GR}$, where $[B_l^m]_{\rm GR}$ denotes the bispectrum multipoles for GR and $[B_l^m]_\gamma$ denotes the multipoles at different values of $\gamma$. The panels from left to right show the results for equilateral ($\mu=0.5,~t=1$), stretched ($\mu=1,~t=0.5$), and squeezed ($\mu=1,~t=1$) triangle shapes, respectively, with fixed size $k_1=0.1$~$\mpc^{-1}$ at $z=0.61$. The black solid line represents monopole ($l=0,m=0$), while different shades of green, blue, and orange-red correspond to $ l=2, 4$ and $6$ multipoles, respectively. Distinct line styles correspond to different $m$ values.}
    \label{fig:gamma_mut}
\end{figure*}

In this subsection, we investigate how the bispectrum multipoles $B_l^m$ respond to modifications in the linear growth of perturbations. Figure~\ref{fig:gamma_mut} shows the ratio $[\alpha_l^m]_\gamma=[B_l^m]_\gamma/[B_l^m]_{\rm GR}$, where $[B_l^m]_{\rm GR}$ denotes the bispectrum multipoles for GR and $[B_l^m]_\gamma$ corresponds to those obtained by varying the linear growth index $\gamma$. The second-order indices are kept fixed at their GR values, i.e., $\psi=\psi_{\rm GR}$ and $\xi=\xi_{\rm GR}$. The results are shown for three distinct triangle configurations $-$ equilateral ($\mu=0.5, t=1$), stretched ($\mu=1, t=0.5$), and squeezed ($\mu=1, t=1$) $-$ shown from left to right across the panels. In the figure, the black solid curve shows the ratio of monopole moments ($l=0$), and the shades of green, blue, and orange‑red respectively show the ratios for quadrupole ($l=2$), hexadecapole ($l=4$), and tetrahexacontapole ($l=6$), as indicated by the colour bar. Different line styles are used to distinguish between multipoles with different $m$ values. The grey solid line shows the ratio of $l=8$ multipole, $[\alpha_8^0]_\gamma$. 

Considering the monopole ratio $[\alpha_0^0]_\gamma$, we see that it is relatively small, around 0.6, at low values of $\gamma$ across all three triangle configurations. It increases steadily with $\gamma$, but remains $\lesssim 1.3$ as $\gamma\rightarrow 1$. Similarly, the quadrupole ($l=2$) ratios $- [\alpha_2^0]_\gamma$, $[\alpha_2^1]_\gamma$, and $[\alpha_2^2]_\gamma -$ also start at lower values for small $\gamma$ and rises with $\gamma$ across all triangle configurations, varying within the range of 0.7 to 1.2. We find that, for certain triangle shapes, all the quadrupole moments carry no independent information on the linear growth index. In equilateral and squeezed configurations, $[\alpha_2^0]_\gamma$ and $[\alpha_2^2]_\gamma$ coincide exactly, while in stretched configurations $[\alpha_2^0]_\gamma$ and $[\alpha_2^1]_\gamma$ are identical for all $\gamma$. Therefore, to fully exploit the potential of all quadrupole moments for probing the gravity, the analysis must include a broader variety of triangle shapes.

Next, we examine the behaviour of the hexadecapole ($l=4$) moments. In general, the ratios $[\alpha_4^m]_\gamma$ for $m = 0, 1, 2, 3, 4$ initially start at a lower value of around 0.8 and typically increase with $\gamma$, reaching a peak close to 1, and subsequently decrease as $\gamma$ continues to grow across all triangle configurations. The transition points, $\gamma = \gamma_t$, where each multipole $[\alpha_4^m]_\gamma$ shifts from rising to falling, vary with the configuration. The transition points $\gamma_t$ for each $m$ multipoles and triangle configuration are summarised in Table~\ref{tab:gamma_t}.

\begin{table}[ht]
\centering
\begin{tabular}{l|ccccc}
\hline
Triangle shape & \multicolumn{5}{c}{$\gamma_t$ values for $l=4$ multipoles} \\
\cline{2-6}
& $m=0$ & $m=1$ & $m=2$ & $m=3$ & $m=4$ \\
\hline
Equilateral  & 0.65 & 0.35 & 0.65 & 0.55 & 0.65 \\
Stretched   & 0.36 & 0.35 & --  & --  & 0.62 \\
Squeezed     & 0.73 & 0.42 & 0.59 & 0.42 & 0.65 \\
\hline
\end{tabular}
\caption{Transition points $\gamma_t$ for the hexadecapole ($l=4$) ratios $[\alpha_4^m]_\gamma$ across different triangle configurations. For stretched triangles, no transition point exists for the $m=2$ and $m=3$ multipoles due to their monotonic decrease with $\gamma$.}
\label{tab:gamma_t}
\end{table}


While most hexadecapole moments exhibit a peak within the range $0.35 \lesssim \gamma_t \lesssim 0.73$, a notable exception arises in the $[\alpha_4^2]\gamma$ and $[\alpha_4^3]\gamma$ components for stretched triangles. In these cases, the ratios decrease monotonically with increasing $\gamma$, starting from approximately 2 as $\gamma \to 0$ and steadily declining to around 0.25 as $\gamma \to 1$. This indicates that these multipoles are more sensitive to variations in $\gamma$. We also observe degeneracies among certain multipoles for specific triangle shapes. For equilateral triangles, we observe that $[\alpha_4^4]_\gamma = [\alpha_4^2]_\gamma = [\alpha_4^0]_\gamma$ throughout the range of $\gamma$, while for stretched triangles, $[\alpha_4^1]_\gamma = [\alpha_4^0]_\gamma$ consistently holds. In the case of squeezed triangles, $[\alpha_4^1]_\gamma$ equals $[\alpha_4^3]_\gamma$ when $\gamma \geq \gamma_{\rm GR}$. As a result, the $m=1, 2, 4$ hexadecapoles carry no additional information on the linear growth index beyond what is already present in the $m=0$ moment for these cases.

We now turn to the behaviour of the tetrahexacontapole ($l=6$) moments.  In general, $[\alpha_6^m]_\gamma$ values are initially high at small $\gamma$ and gradually decrease as $\gamma$ increases. However, the $[\alpha_6^1]_\gamma$ corresponding to equilateral triangles is an exception, which begins at a lower value and surges as $\gamma$ becomes large. Additionally, the ratios $[\alpha_6^0]_\gamma$, $[\alpha_6^1]_\gamma$, $[\alpha_6^2]_\gamma$, and $[\alpha_6^3]_\gamma$ also exhibit a distinct behaviour for the case of squeezed triangles. Their values are in the range $1-1.1$ at small $\gamma$, and they increase slightly to attain a peak value at around $\gamma_t=0.17$ and then begin to decline gradually. The multipoles $[\alpha_6^0]_\gamma$, $[\alpha_6^2]_\gamma$ and $[\alpha_6^4]_\gamma$ for equilateral configuration are especially sensitive to the $\gamma$. They attain values in the range $\approx 2-2.6$ at small $\gamma\rightarrow 0$ and decline to $0-0.4$ at $\gamma=1$. For the rest of the cases,  $[\alpha_6^m]_\gamma$ remains within the narrower range from $0.5$ to $1.4$. In many cases, various $[\alpha_6^m]_\gamma$ overlap with each other, and not all tetrahexacontapoles contain independent information on linear growth. For the case of stretched triangles, $[\alpha_6^0]_\gamma=[\alpha_6^1]_\gamma$ and $[\alpha_6^2]_\gamma\approx[\alpha_6^3]_\gamma\approx[\alpha_6^5]_\gamma$. For squeezed triangles, $[\alpha_6^0]_\gamma\approx[\alpha_6^1]_\gamma\approx[\alpha_6^2]_\gamma$ and $[\alpha_6^4]_\gamma=[\alpha_6^5]_\gamma=[\alpha_6^6]_\gamma$. 

Finally, we consider the highest non-zero multipoles, $l=8$, where only terms with $m\leq6$ are non-vanishing, while all higher-order multipoles are identically zero. We see that these $l=8$ multipoles do not provide additional information about $\gamma$ beyond what is already captured by the $l=6$ multipoles. For all triangle configurations, each $[\alpha_8^m]_\gamma$ multipole is precisely equal to $[\alpha_6^6]_\gamma$.


\begin{figure*}
    \centering 
    \includegraphics[width=.99\textwidth]{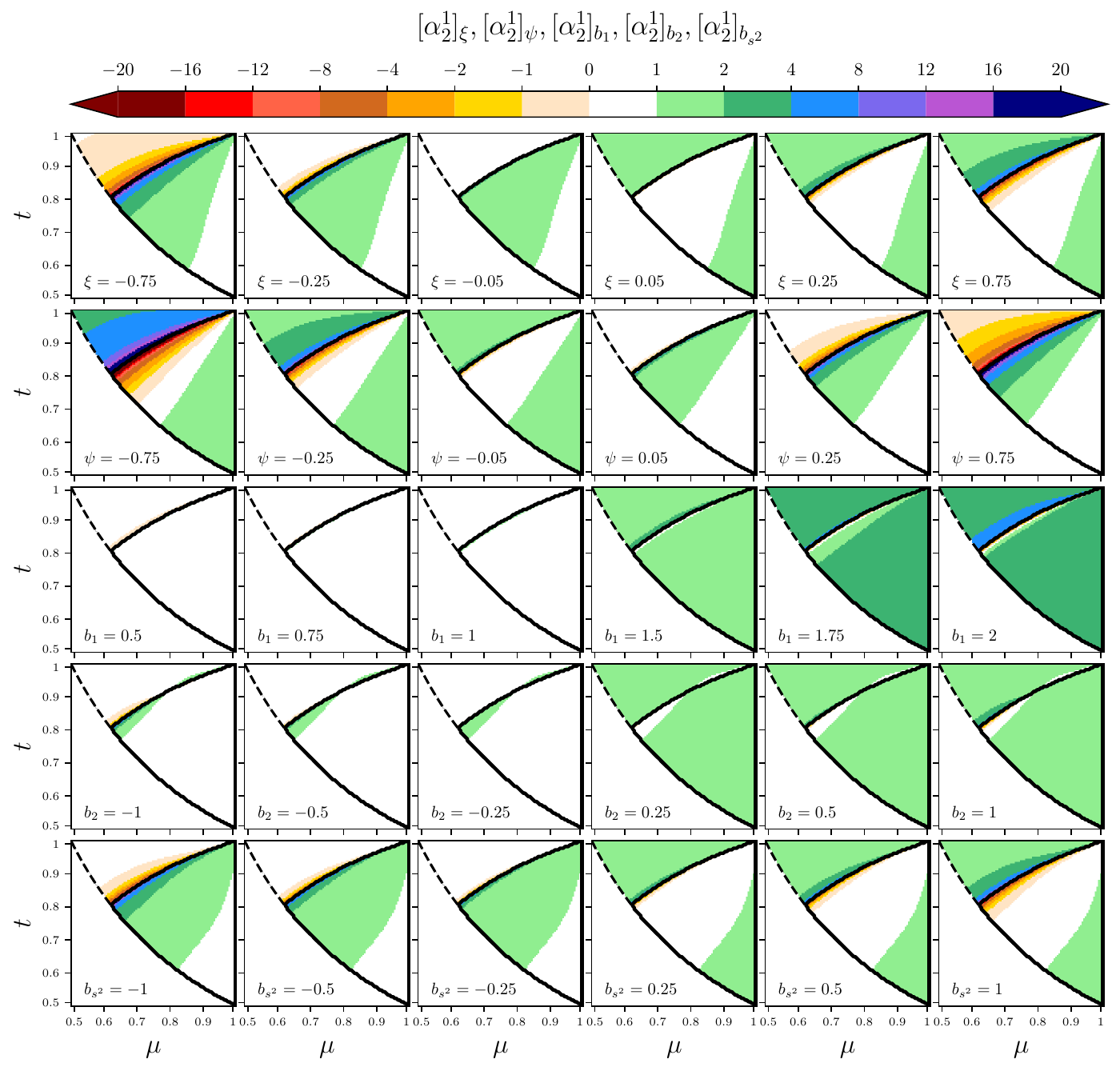} 
    \caption{Sensitivity of $B_2^1$ to variations in the second-order growth indices $\xi$ and $\psi$. The first and second rows show the ratio $[\alpha_2^1]_\xi=[B_2^1]_\xi/[B_2^1]_{\rm GR}$ and $[\alpha_2^1]_\psi=[B_2^1]_\psi/[B_2^1]_{\rm GR}$, respectively, across all possible unique triangle configurations. Here $[B_2^1]_\xi$ and $[B_2^1]_\psi$ are the bispectrum multipoles at different values of $\xi$ and $\psi$ respectively, while $[B_2^1]_{\rm GR}$ corresponds to the values considering the GR. Different columns correspond to specific values of $\xi$ and $\psi$, as specified in the respective panels. Each subplot illustrates results in $\mu-t$ space, where each point corresponds to a distinct triangle shape (refer to Figure 2 of \citep{Bharadwaj:2020MNRAS.493..594B}), with triangle size fixed at $k_1=0.1$ $\mpc^{-1}$ fixed. The third, fourth, and fifth rows show the $[\alpha_2^1]_{b_1}$, $[\alpha_2^1]_{b_2}$, and $[\alpha_2^1]_{b_{s^2}}$, respectively, depicting the impact of linear and quadratic bias parameters on the bispectrum multipoles. The solid black contours highlight the locations where $[B_2^1]_{\rm GR}$ has a negative sign.}
    \label{fig:b21_mut}
\end{figure*}

\subsection{Sensitivity to second-order growth}
Further, we analyse the sensitivity of bispectrum multipoles $B_l^m(k_1,\mu,t)$ to variations in the second-order growth indices, $\xi$ and $\psi$. The observed $B_l^m$ is also influenced by the bias of tracers (Equation~\ref{eq:final_bs}); therefore, quantifying the effects due to the bias is crucial to isolating the true signal indicative of deviations from GR. We study the variations of $B_l^m$ with respect to both the second-order growth indices and bias parameters simultaneously. 

To thoroughly assess the capability of $B_l^m(k_1,\mu,t)$ in distinguishing second‑order growth effects induced by modified gravity, we systematically explore all possible triangle configurations in our analysis. Here, the parameters $(\mu, t)$ uniquely specify the shape of each triangle.  Thus, to investigate the behaviour of $B_l^m(k_1, \mu, t)$ across the full range of triangle shapes, we present our results in the $\mu$–$t$ parameter space while keeping the triangle size fixed at $k_1 = 0.1~\mathrm{Mpc}^{-1}$. Here, we focus our discussion on the multipoles that exhibit significant indications of modified gravity, while the remaining multipoles are discussed in the \ref{sec:appendix}. Notably, the monopole $B_0^0$, as shown in Figure~\ref{fig:b00}, exhibits minimal variation with changes in gravity at the second order, and its corresponding results are provided in the \ref{sec:appendix}. The results for the higher-order multipoles are discussed in detail below.

\subsubsection{Quadrupole ($l=2$) moments}

Figure~\ref{fig:b21_mut} shows the results for $l=2,m=1$ multipole. In the top row, we plot the ratio $[\alpha_l^m]_\xi=[B_l^m]_\xi/[B_l^m]_{\rm GR}$ as a function of $(\mu,t)$, with the $k_1=0.1$ Mpc$^{-1}$ fixed. Here, $[B_l^m]_\xi$ denotes the bispectrum multipoles evaluated by varying $\xi$, while the other growth indices are held fixed at their GR values, i.e., $\gamma=\gamma_{\rm GR}$ and $\psi=\psi_{\rm GR}$. Subsequent rows (second to fifth) present the ratios $[\alpha_l^m]_\psi$, $[\alpha_l^m]_{b_1}$, $[\alpha_l^m]_{b_2}$ and $[\alpha_l^m]_{b_{s^2}}$, respectively, each defined analogously to $[\alpha_l^m]_\xi$. The results for six distinct values of these parameters ($\xi,\psi,b_1,b_2,b_{s^2}$) are presented in each column.

Each subplot in Figure~\ref{fig:b21_mut} shows results across the entire allowed range of $\mu$ and $t$ values, bounded by $0.5 \leq \mu, t \leq 1$ and $2\mu t \geq 1$. This region covers triangles of all possible shapes, with each point $(\mu, t)$ representing a unique triangle configuration. For a detailed discussion on the locations of various triangle shapes in the $\mu-t$ plane, refer to Figure 2 in \citep{Bharadwaj:2020MNRAS.493..594B}. Briefly, the upper left corner ($\mu=0.5, t=1$) represents the equilateral triangles. The upper and lower boundaries, marked by $t=1$ and $2\mu t=1$, respectively, depict L isosceles triangles (the two largest sides are equal; $t=1$) and S isosceles triangles (the two smallest sides are equal; $t=0.5/\mu$). The right boundary ($\mu=1$) represents linear triangles, where the three sides are nearly collinear. Furthermore, the upper right ($\mu=1,t=1$) and lower right ($\mu=1,t=0.5$) corners correspond to the squeezed and stretched triangles. The relation $\mu=t$ corresponds to the right-angle triangles, which separates the regions with acute triangles ($\mu < t$) from those of obtuse triangles ($\mu > t$). 


Considering the first row of Figure~\ref{fig:b21_mut}, we observe that the magnitude of the ratio $|[\alpha_2^1]_\xi|$ remains below 8 across most of the $\mu-t$ plane for all values of $\xi$. Generally, the ratio stays close to unity at the stretched limit ($\mu=1,t=0.5$), and it increases as triangles are deformed from obtuse ($\mu>t$) to acute ($\mu<t$). The peak value lies near the right-angle triangles. As the triangle shapes are further deformed toward the equilateral limit, $|[\alpha_2^1]_\xi|$ begins to decline and approaches a value close to unity. In particular, for the small $\xi\in [-0.05,0.05]$, the ratio $|[\alpha_2^1]_\xi|$ remains in the order of unity for all triangles, indicating that the minor variations in $\xi$ have little effect on the $B_2^1$ multipole. 
When $\xi=\pm 0.25$, a few acute triangles near right‑angle limit see the ratio rise to around 4. At the extreme values $\xi=\pm0.75$, the $|[\alpha_{2}^{1}]_{\xi}|$ for these acute configurations reaches between 4 and 12. Thus, the $B_2^1$ multipole exhibits the highest sensitivity to $\xi$ for these acute triangle configurations, suggesting that such triangles provide an optimal window to probe $\xi$ using $B_2^1$.

Considering next the $[\alpha_2^1]_\psi$ shown in the second row, we see that the results for $\psi<0$ closely resemble those for $\xi>0$ and vice versa. The overall magnitude of ratio $[\alpha_2^1]_\psi$ is greater than that of $[\alpha_2^1]_\xi$ for the identical values of the parameters $\psi$ and $\xi$. For most cases, the $|[\alpha_2^1]_\psi|$ remains below $12$. However, for $\psi=-0.75$, the $|[\alpha_2^1]_\psi|$ has values in the range $16-20$ for a few acute triangles, and the values decline towards the equilateral limit to $\sim 2-4$. The $[\alpha_2^1]_{\psi=-0.75}\sim 0-2$ for all obtuse triangles. The $|[\alpha_2^1]_\psi|$ declines as $\psi$ is increased towards $\psi_{\rm GR}$. At $\psi=\pm 0.05$, $[\alpha_2^1]_\psi$ remains between $0-2$ for all triangles. As $\psi$ is increased further to $0.75$, the $[\alpha_2^1]_\psi$  begin to surge and reached $\sim 20$ near right-angle limit.

In addition to the variation in magnitude, the sign of $B_2^1$ multipole also serves as a crucial indicator of deviation from GR. We find that the ratios $[\alpha_2^1]_\xi$ and $[\alpha_2^1]_\psi$ becomes negative for certain acute triangle configurations, implying that $[B_2^1]_\xi$ and $[B_2^1]_\psi$ have the opposite signs to that of $[B_2^1]_{\rm GR}$. This observation suggests that if galaxy surveys measure the $B_2^1$ with a sign contrary to GR predictions, it provides clear evidence of a departure from GR. Moreover, the sign of $B_2^1$ not only signals a deviation from GR but also distinguishes between the scenarios of $\xi<0$ (or $\psi>0$) and $\xi>0$ (or $\psi<0$). The specific triangle configuration for which the sign of $B_2^1$ flips provides key insight into the sign of second-order indices $\xi$ and $\psi$.

Here, we have identified the triangle shapes that exhibit this sign-flip signature for both $\xi<0$ (equivalently $\psi>0$) and $\xi>0$ (equivalently $\psi<0$). 
In the plots, we use solid black contours to demarcate the regions where the GR predicts a negative value for the $B_2^1$ multipole, i.e., where $[B_2^1]_{\rm GR}<0$. The contours span all obtuse triangle configurations ($\mu > t$) and extend slightly into the acute regime near the right-angle limit ($\mu \approx t$).  For $\xi < 0$ (or $\psi>0$), we observe that $[\alpha_2^1]_\xi$ (or $[\alpha_2^1]_\psi$) becomes negative for acute triangles close to the equilateral limit. This indicates that $[B_2^1]_\xi$ (or $[B_2^1]_\psi$) is negative for these triangle configurations where GR predicts a positive value $[B_2^1]_{\rm GR}>0$, providing a signature of modification to GR with $\xi < 0$ or $\psi>0$. Conversely, for $\xi > 0$ (or $\psi<0$), the ratio $[\alpha_2^1]_\xi$ (or $[\alpha_2^1]_\psi$) becomes negative near the right-angle limit, where $[B_2^1]_{\rm GR}$ is negative. This implies that $[B_2^1]_\xi$ (or $[B_2^1]_\psi$) has flipped to a positive value, offering a distinct indicator of deviation from GR associated with $\xi > 0$ or $\psi>0$. Notably, the number of triangle configurations exhibiting this sign flip increases with the strength of the deviation, i.e., with larger values of $|\xi|$ and $|\psi|$. Specifically, for $\psi=0.75$, the multipole $[B_2^1]_{\psi}$ is negative across all triangle configurations, in contrast to the GR prediction where $[B_2^1]_{\rm GR}$ remains positive for many configurations near the equilateral limit. In a nutshell, the $B_2^1$ exhibits significant variations with $\xi$ and $\psi$ across most triangle configurations, and its negative sign for equilateral triangles and positive values near right-angle triangles serves as a remarkable signature of modified gravity.

The bias parameters can also affect the $B_2^1$ and may potentially produce false signals of modified gravity if the bias of tracers is not accurately known a priori. Further, we analyse how linear bias $b_1$ and quadratic bias parameters $b_2, b_{s^2}$ affect the $B_2^1$. Third row in Figure~\ref{fig:b21_mut} presents ratio $[\alpha_2^1]_{b_1}$. We see that for small values of $b_1$ in the range 0.5 to 1, the $[\alpha_2^1]_{b_1}$ is less than 1 for all triangles. For large values where $b_1=1.5$ and $1.75$, the $[\alpha_2^1]_{b_1}$ remains within the range $1-2$ and $2-4$ respectively. The $[\alpha_2^1]_{b_1=2}$ is also within the 2-4 range for most triangles, except for a few acute triangles where it increases to approximately  $\sim 4-8$. Considering the fourth and fifth rows, we see that the $|[\alpha_2^1]_{b_2}|$ and $|[\alpha_2^1]_{b_{s^2}}|$ remain below $2$ for most of the triangles. The ratio $|[\alpha_2^1]_{b{s^2}}|$ becomes negative for a few acute triangle configurations near the right-angle limit, suggesting that a large offset in the parameter ${b_{s^2}}$ can potentially mimic a false signal of modified gravity. However, within the range of all bias parameters considered in this analysis, we do not observe any sign flip in $B_2^1$ near the equilateral limit—a behaviour that is characteristic of modified gravity.

This concludes our discussion of the $B_2^1$ multipole. Results for the other quadrupole moments—$B_2^0$ and $B_2^2$—are presented in \ref{sec:appendix}.



\subsubsection{Hexadecapole ($l=4$) moments}

\begin{figure*}
    \centering 
    \includegraphics[width=.99\textwidth]{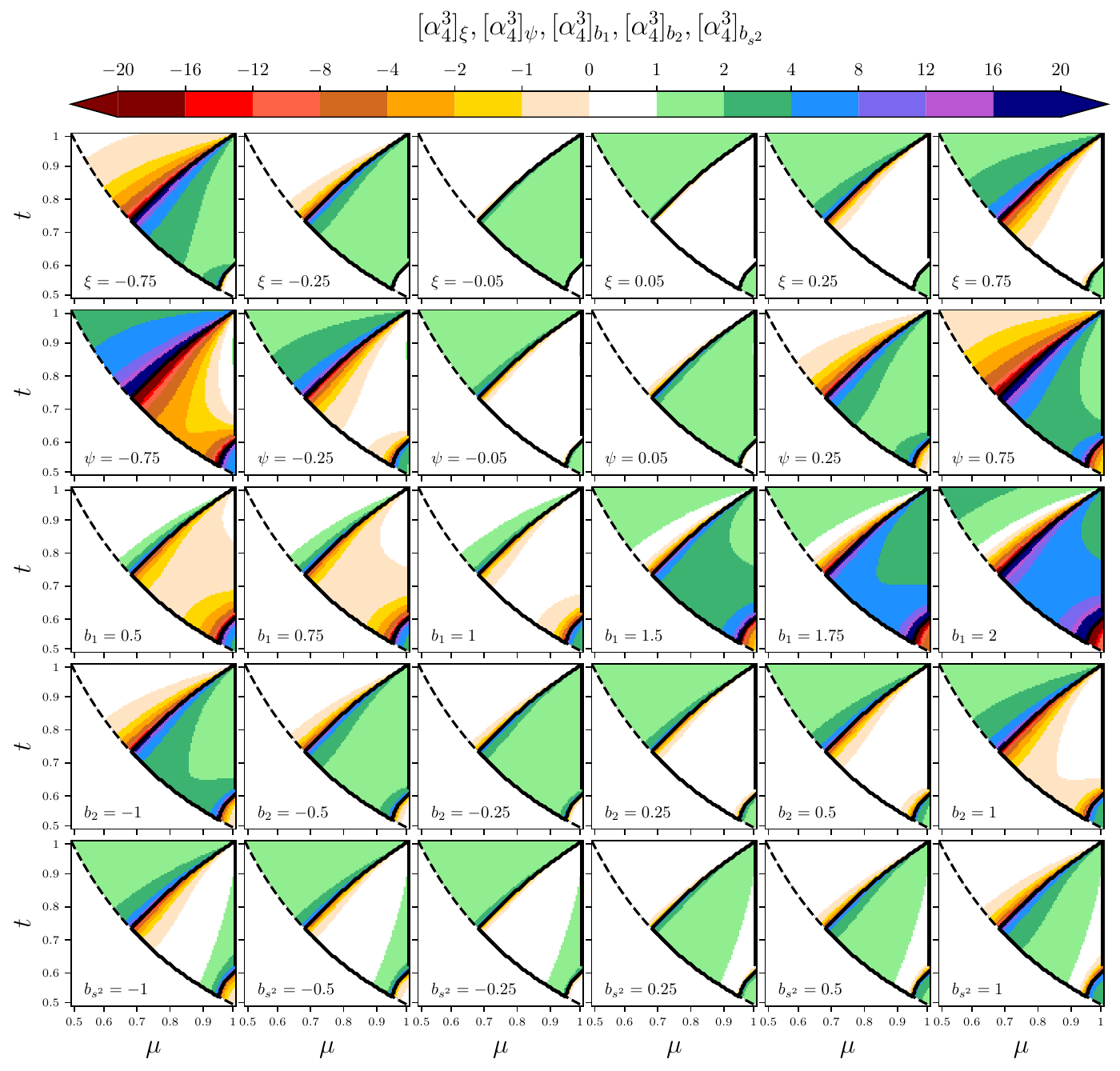} 
    \caption{Sensitivity of $B_4^3$ to variations in the second-order growth indices $\xi$ and $\psi$. The rows, from top to bottom, correspond to $[\alpha_4^3]_{\xi}$, $[\alpha_4^3]_{\psi}$, $[\alpha_4^3]_{b_1}$, $[\alpha_4^3]_{b_2}$, and $[\alpha_4^3]_{b_{s^2}}$, respectively.  The panels are identical to those in Figure~\ref{fig:b21_mut}.}
    \label{fig:b43_mut}
\end{figure*}


Next, we discuss the hexadecapole ($l=4$) moments. The results for the $l = 4, m = 3$ multipole are shown in Figure~\ref{fig:b43_mut}, while the corresponding results for the other $l = 4$ multipoles are provided in \ref{sec:appendix}. Considering the first row, we see that the trend of the ratio $[\alpha_4^3]_\xi$ in the $\mu-t$ plane is somewhat similar to the $[\alpha_2^1]_\xi$ case. The notable difference is that the magnitude of the ratio $|[\alpha_4^3]_\xi|$ is largest near the right-angle configurations ($\mu = t$), reaching values as high as $\sim 20$ for $\xi = \pm 0.75$. This amplitude gradually decreases as the triangle shape deviates from the right-angle limit, becoming either more acute ($\mu < t$) or more obtuse ($\mu > t$). Nonetheless, $|[\alpha_4^3]_\xi|$ exhibits a slight increase near the most obtuse configurations—close to the stretched limit—with values rising to approximately $\sim 1$–$2$ for $\xi = 0.75$ and $\sim 2$–$4$ for $\xi = -0.75$. For smaller values of $\xi$, the peak values near the right-angle configuration are reduced, with $|[\alpha_4^3]_{\xi = \pm 0.25}| \sim 10$ and $|[\alpha_4^3]_{\xi = \pm 0.05}| \sim 3$.

We now consider $[\alpha_4^3]_\psi$, shown in the second row.  Much like the quadrupole case, the behaviour observed in the $\mu-t$ space reflects a pattern similar to that when $\psi \rightarrow -\xi$. The overall trend of $[\alpha_4^3]_\psi$ closely resembles that of $[\alpha_4^3]_\xi$, with the magnitude $|[\alpha_4^3]_\psi|$ showing a prominent peak near right-angle configurations. However, the magnitude of $|[\alpha_4^3]_\psi|$ is generally larger than that of $|[\alpha_4^3]_\xi|$ across the shape space. At the right-angle limit, the peak values of $|[\alpha_4^3]_\psi|$ reach approximately $20$, $16$, and $12$ for $\psi = \pm 0.75$, $\pm 0.25$, and $\pm 0.05$, respectively. Beyond this pronounced peak, $|[\alpha_4^3]_\psi|$ exhibits a secondary rise in amplitude near the stretched limit ($\mu \rightarrow 1$, $t \rightarrow 0.5$), particularly notable for higher values of $\psi = \pm 0.75$, where $|[\alpha_4^3]_\psi|$ reaches up to $16$.

Similar to the quadrupole case, the $B_4^3$ multipole also exhibits sign reversals across a range of triangle configurations. Notably, this effect is more prominent here, with a greater number of triangles exhibiting sign flips near the right-angle ($\mu = t$) and stretched limit ($\mu \rightarrow 1$, $t \rightarrow 0.5$). Under GR, we see that the $[B_4^3]_{\rm GR}$ is positive for acute triangles and negative for obtuse triangles, except for a few triangles near the stretched limit. However, $[B_4^3]_\xi$ for $\xi < 0$ and  $[B_4^3]_\psi$ for $\psi > 0$ become negative for many acute triangles—regions where GR predicts a positive value. Conversely, $[B_4^3]_\xi$ for $\xi > 0$ and $[B_4^3]_\psi$ for $\psi < 0$, turn positive for many obtuse triangles where GR predicts a negative value. This clear contrast in the sign of $B_4^3$ between GR and the DHOST scenario serves as a strong indicator of modifications in gravity.

Nonetheless, the sign flip may also be induced by the galaxy bias parameters, as discussed earlier. The third, fourth and fifth rows in the Figure~\ref{fig:b43_mut} shows $[\alpha_4^3]_{b_1}$, $[\alpha_4^3]_{b_2}$ and $[\alpha_4^3]_{b_{s^2}}$ respectively. We see that the impact of considering different $b_1$, $b_2$, and $b_{s^2}$ parameters is severe in the case of $l=4,m=3$ moment. The $[\alpha_4^3]_{b_1}$ is small for acute triangles near the equilateral limit, where its magnitude is $< 1$ if $b_1<1.2$, and it is around $1-4$ if $b_1>1.2$. However, the $[\alpha_4^3]_{b_1}$ is highest ($\sim 16$ for $b_1<1.2$ and $\sim 20$ for $b_1>1.2$) near the right-angle and stretched limit. Furthermore, we observe that $[B_4^3]_{b_1}$ changes sign for many obtuse triangles when $b_1 < 1.2$ and for a few acute and stretched triangles when $b_1 > 1.2$. This behaviour closely resembles the sign flip of $[\alpha_4^3]_\psi$. Similarly, the influence of second-order bias parameters is most pronounced near the right-angle and stretched configurations, while it is relatively small ($|[\alpha_4^3]_{b_2}| \approx |[\alpha_4^3]_{b_{s^2}}| \sim 1-2$) near equilateral and some obtuse triangle configurations. Notably, the trend of $[\alpha_4^3]_{b_2}$ closely follows that of $[\alpha_4^3]_{\xi}$. In the case of $b_{s^2}$, the $[\alpha_4^3]_{b_{s^2}}$ changes sign for a few obtuse triangles when $b_{s^2} < 0$ and for a few acute triangles when $b_{s^2} > 0$, similar to $\psi$. However, it does not fully mimic $\psi$, as it does not change the sign above the lower boundary of the black curve for $b_{s^2} < 0$ or near the stretched limit for $b_{s^2} > 0$. Instead, for $b_{s^2} < 0$, it changes the sign near the stretched limit, and for $b_{s^2} > 0$, it changes the sign above the lower boundary of the black curve, which are the features of the $\xi$ parameter. Thus, $b_{s^2}$ exhibits characteristics of both the $\xi$ and $\psi$ parameters. Consequently, these bias parameters must be accurately accounted for to utilise $B_4^3$ as a probe for the DHOST theories effectively.

\subsubsection{Tetrahexacontapole ($l=6$) moments}

\begin{figure*}
    \centering 
    \includegraphics[width=.99\textwidth]{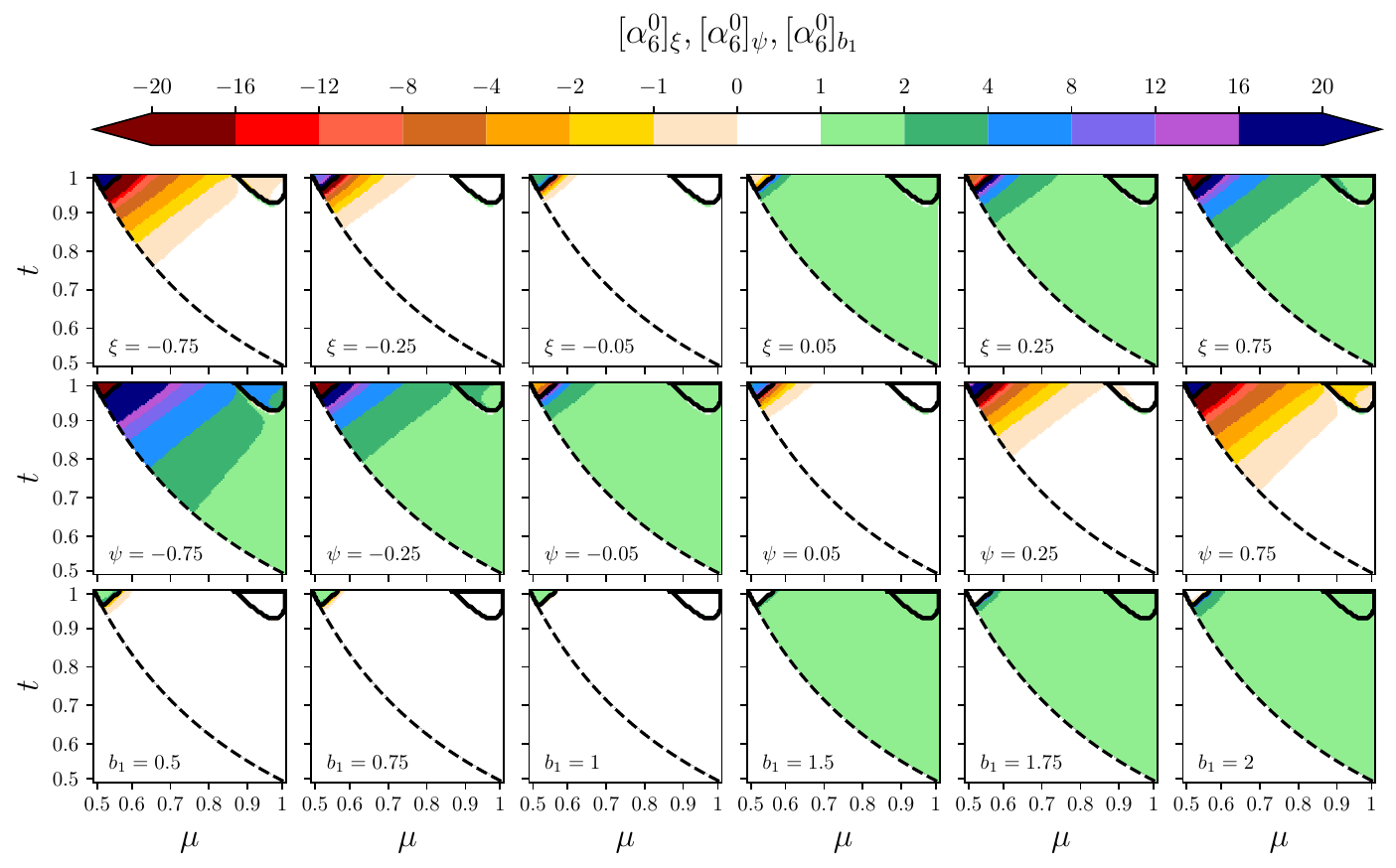} 
    \caption{Sensitivity of $B_6^0$ to variations in the second-order growth indices $\xi$ and $\psi$. The top, middle, and bottom rows show $[\alpha_6^0]_{\xi}$, $[\alpha_6^0]_{\psi}$, and $[\alpha_6^0]_{b_1}$ respectively. The panels are the same as in Figure~\ref{fig:b21_mut}, except for the last two columns, as $B_6^0$ is unaffected by quadratic bias and second-order tidal bias parameters.}
    \label{fig:b60_mut}
\end{figure*}

We now turn to the analysis of the $l=6$ multipoles. Beginning with the case of $l=6, m=0$, illustrated in Figure~\ref{fig:b60_mut}, we find that the $B_6^0$ multipole exhibits high sensitivity to variations in the parameters $\xi$ and $\psi$. This sensitivity is most pronounced for acute configurations $(\mu<t)$, particularly near the equilateral limit, where $B_6^0$ demonstrates substantial changes even for small $\xi$ and $\psi$. Specifically, the maximum value of $|[\alpha_6^0]_\xi|$ reaches approximately $\sim 8$ for $\xi = \pm 0.05$, while $|[\alpha_6^0]_\psi|$ rises steeply to attain values as high as $\sim 12$ for $\psi = \pm 0.05$ near the equilateral limit. However, for other triangle configurations, the magnitudes of the ratios remain relatively small: $[\alpha_6^{0}]_{\xi = -0.05} \approx [\alpha_6^{0}]_{\psi = 0.05} \sim 1$ and $[\alpha_6^{0}]_{\xi = 0.05} \approx [\alpha_6^{0}]_{\psi = -0.05} \sim 2$. The values of $|[\alpha_6^0]_\xi|$ and $|[\alpha_6^0]_\psi|$ increase gradually with higher values of $\xi$ and $\psi$, respectively. For equilateral triangles, both $|[\alpha_6^0]_\xi|$ and $|[\alpha_6^0]_\psi|$ exceed $20$ when $\xi$ and $\psi$ attain values $\pm 0.75$. The number of triangle configurations where both $|[\alpha_6^0]_\xi|$ and $|[\alpha_6^0]_\psi|$ exceed $1$ increases significantly in the acute triangle region as $\xi$ and $\psi$ upsurge. In contrast, for the obtuse triangles, $[\alpha_6^0]_\xi$ and $[\alpha_6^0]_\psi$ remains largely below $2$. 

Similar to the cases of both the quadrupole and hexadecapole moments, the sign of $B_6^0$ also serves as a valuable indicator of deviations from GR. The $[B_6^0]_{\text{GR}}$ is negative for a few triangles near the equilateral and squeezed limit. We see that the $[\alpha_6^0]_{\xi<0}$ and $[\alpha_6^0]_{\psi>0}$ are negative for many triangles in the region below the black contour near the equilateral limit, while $[\alpha_6^0]_{\xi>0}$ and $[\alpha_6^0]_{\psi<0}$ are negative for equilateral triangles. This implies that a measured positive $B_6^0$ for equilateral triangles or a negative $B_6^0$ for acute triangles (below black contour) serves as a definitive indicator of deviations from GR. 

As noted earlier for the $l=2$ and $l=4$ multipoles, the apparent signal of modified gravity in $B_6^0$ can be induced by the bias parameters. However, the $l=6$ multipoles offer a distinct advantage over the $l=2$ and $l=4$ multipoles as they are unaffected by the quadratic bias $b_2$ and second-order tidal bias $b_{s^2}$ parameters. Considering Equation~\ref{eq:final_bs}, we note that the terms involving $b_2$ and $b_{s^2}$ are associated with the $S_{12}$ term, which is a polynomial in $\mu_i^2$ with the highest power term being the $\mu_i^4$. So, bispectrum multipoles with $l\leq 4$ are affected by the higher-order bias terms. Only the linear bias $b_1$ influences the $l=6$ multipoles. The third row of the Figure~\ref{fig:b60_mut} shows $[\alpha_6^0]_{b_1}$. We see that, for $b_1<1.2$, the $|[\alpha_6^0]_{b_1}|<1$ for most of the triangle configurations except at the equilateral limit where it has value $1-2$. For $b_1>1.2$, $|[\alpha_6^0]_{b_1}|\sim 1-2$. We see that the small values of $b_1$ can induce the negative sign in the $B_6^0$ below the equilateral limit. Overall, the influence of bias parameters on $B_6^0$ is relatively minor and does not pose a significant concern. 


\begin{figure*}
    \centering 
    \includegraphics[width=.99\textwidth]{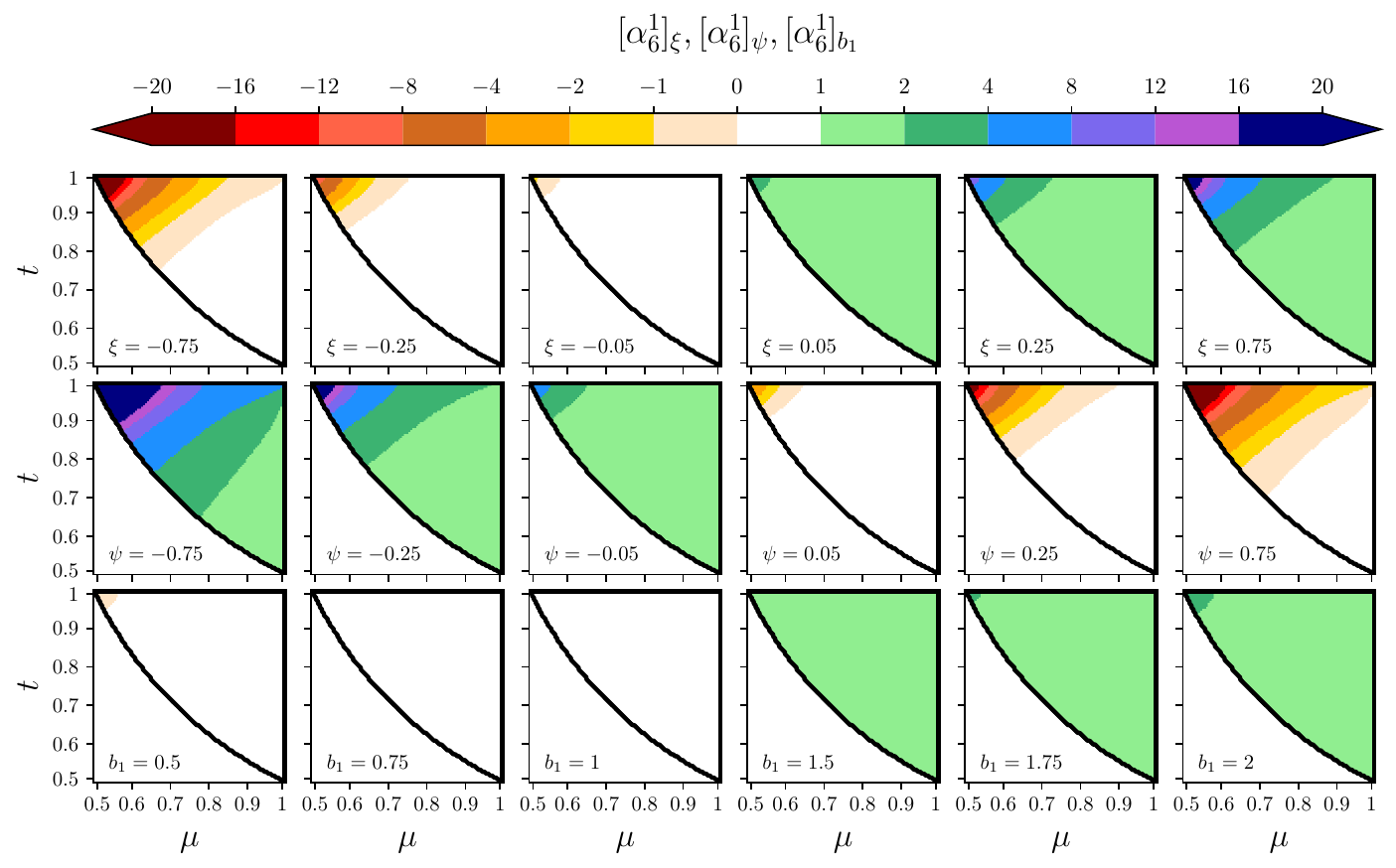} 
    \caption{Sensitivity of $B_6^1$ to variations in the second-order growth indices $\xi$ and $\psi$. The top, middle, and bottom rows correspond to $[\alpha_6^1]_{\xi}$, $[\alpha_6^1]_{\psi}$, and $[\alpha_6^1]_{b_1}$ respectively. The panels are identical to those in Figure~\ref{fig:b60_mut}.}
    \label{fig:b61_mut}
\end{figure*}
Figure~\ref{fig:b61_mut} shows the results for $l=6,m=1$ multipole. We see that the results are very similar to the $l=6,m=0$ multipole. The $|[\alpha_6^{1}]_\xi|$ and $|[\alpha_6^{1}]_\psi|$ are large for acute triangles, especially the equilateral limit, which is sensitive to very small values of $\xi$ and $\psi$. In contrast to $m=0$ case, the $[B_6^1]_{\text{GR}}$ is negative for all triangle configurations. We see that the $[\alpha_6^{1}]_{\xi>0}$ and $[\alpha_6^{1}]_{\psi<0}$ are positive for all triangles, and $[\alpha_6^{1}]_{\xi<0}$ and $[\alpha_6^{1}]_{\psi>0}$ are negative for many acute triangles. Thus, the positive values of the observed $B_6^1$ for acute triangles serve as indicators of modifications in GR. The third row shows that the linear bias parameter has a small impact on $B_6^1$ and $[\alpha_6^1]_{b_1}< 1$ for all triangles when $b_1=0.5-1.2$ and $[\alpha_6^1]_{b_1}\sim 1-2$ for most of the triangles (with the exception of equilateral triangles, where $[\alpha_6^1]_{b_1}\sim 2-4$) when $b_1=1.2-2$. Similar to the $m=0$ case, for small values of $b_1$, $[\alpha_6^1]_{b_1}$ is negative near the equilateral limit.

\begin{figure*}
    \centering 
    \includegraphics[width=.99\textwidth]{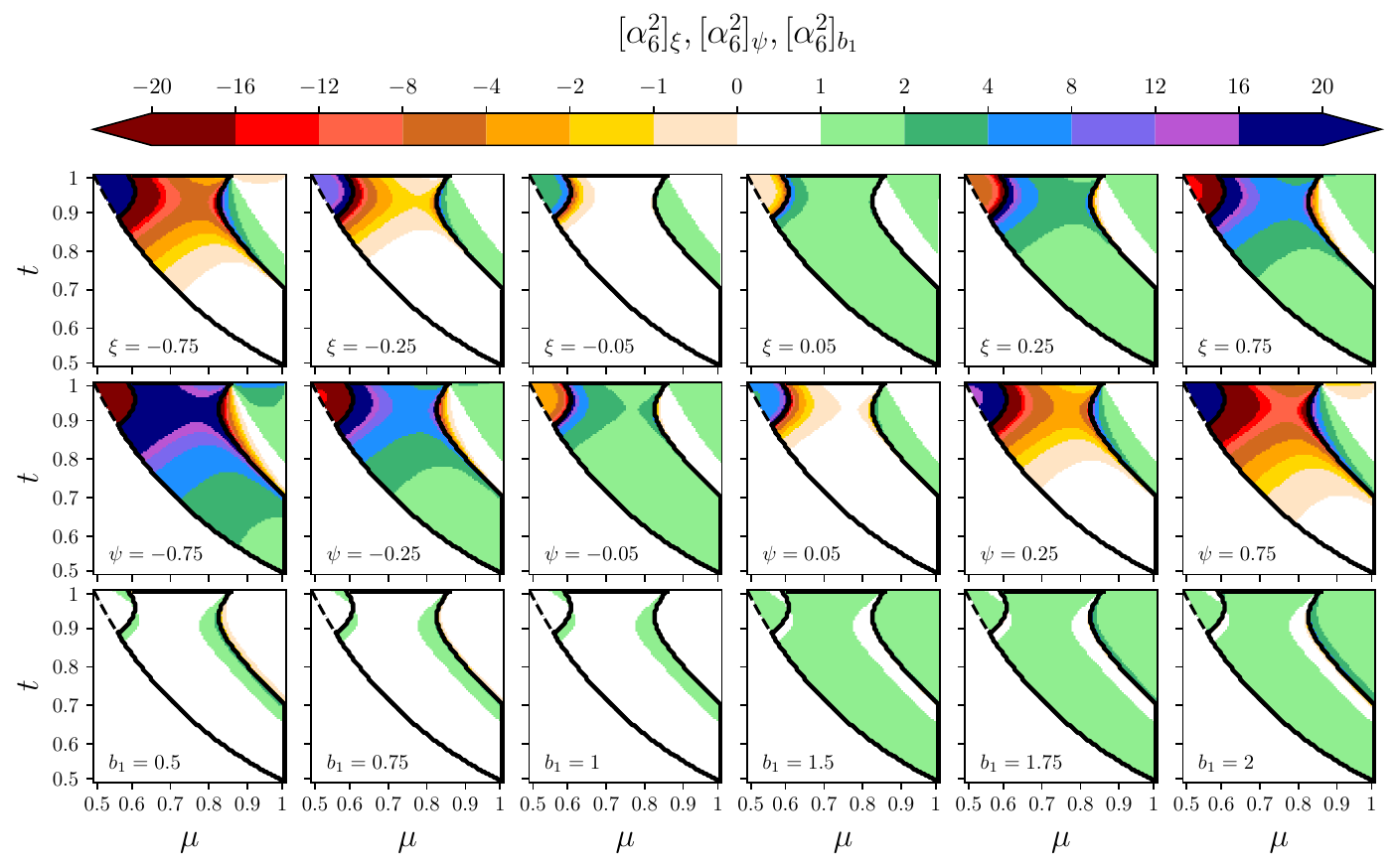} 
    \caption{Sensitivity of $B_6^2$ to variations in the second-order growth indices $\xi$ and $\psi$. The top, middle, and bottom rows correspond to $[\alpha_6^2]_{\xi}$, $[\alpha_6^2]_{\psi}$, and $[\alpha_6^2]_{b_1}$ respectively. The panels are identical to those in Figure~\ref{fig:b60_mut}.}
    \label{fig:b62_mut}
\end{figure*}
Figure~\ref{fig:b62_mut} shows that the  $l=6,m=2$ multipole exhibits significant variation with $\xi$ and $\psi$. The $[\alpha_6^{2}]_{\xi=-0.75}$ and $[\alpha_6^{2}]_{\psi=0.75}$ exceed $20$ for equilateral triangles. As triangles deformed away from the equilateral triangles, $[\alpha_6^{2}]_{\xi=-0.75}$ and and $[\alpha_6^{2}]_{\psi=0.75}$ becomes $\sim -20$. The magnitude of both decreases as we move from acute $(\mu<t)$ to obtuse $(\mu>t)$ triangles. In the region bounded by black contour, the  $[B_6^2]_{\rm GR}$ is negative, however, $[B_6^2]_{\xi=-0.75}$ and $[B_6^2]_{\psi=0.75}$ is positive for all the acute triangles, which indicated the deviation from GR. As $\xi$ increases towards $-0.05$ (or $\psi$ is decreased to 0.05), the overall $[\alpha_6^{2}]_\xi$ (or $[\alpha_6^{2}]_\psi$) declines, however its magnitude remains between $2-8$ at equilateral limit. For positive values of $\xi$ (or negative values of $\psi$), $[\alpha_6^{2}]_\xi$ (or $[\alpha_6^{2}]_\psi$) becomes negative for the many triangles near the equilateral limit, and its magnitude remains similar to the case of negative $\xi$ values. Considering the third row, we observe that the $[\alpha_6^2]_{b_1}<2$ across all cases in the range $b_1=0.5-2$. An additional significant advantage of the $B_6^2$ multipole is that the sign of $[\alpha_6^2]_{b_1}$ remains unaffected with $b_1$ in the region where $\xi$ and $\psi$ induce the negative sign. Thus, the negative $B_6^2$ at equilateral and the positive $B_6^2$ at many acute triangles provide a robust signal for modified gravity.

\begin{figure*}
    \centering 
    \includegraphics[width=.99\textwidth]{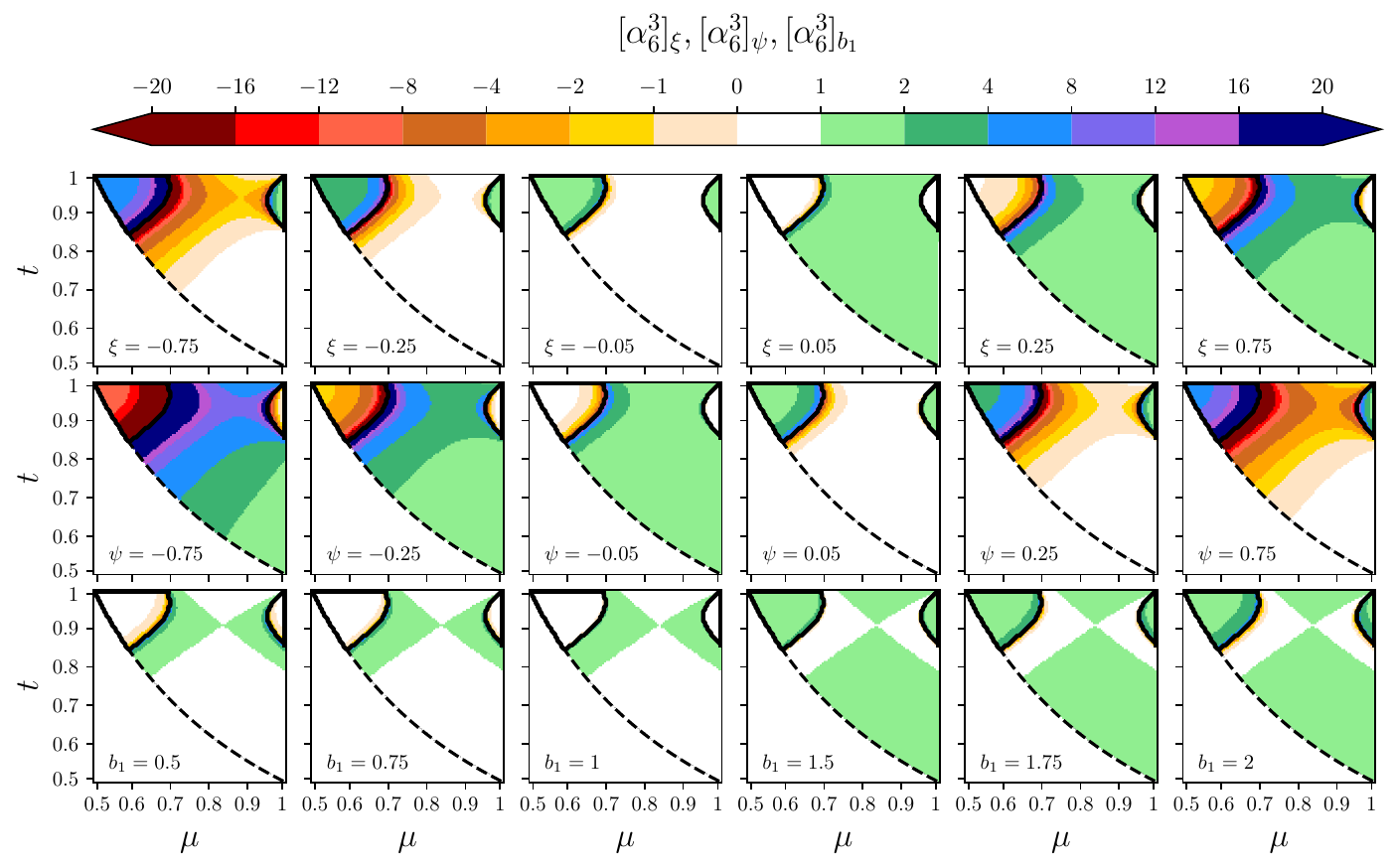} 
    \caption{Sensitivity of $B_6^3$ to variations in the second-order growth indices $\xi$ and $\psi$. The top, middle, and bottom rows correspond to $[\alpha_6^3]_{\xi}$, $[\alpha_6^3]_{\psi}$, and $[\alpha_6^3]_{b_1}$ respectively. The panels are identical to those in Figure~\ref{fig:b60_mut}.}
    \label{fig:b63_mut}
\end{figure*}
 Further, considering Figure~\ref{fig:b63_mut}, we see that the $l=6,m=3$ case is very similar to $m=2$, with the exception that $[B_6^3]_{\rm GR}$ is negative for equilateral triangles and a few linear triangles near the squeezed limit. Similar to $m=2$, $[\alpha_6^{3}]_{\xi}$ and  $[\alpha_6^{3}]_{\psi}$ is negative corresponding to a few acute triangles. The magnitude $|[\alpha_6^{3}]_\xi|$ and $|[\alpha_6^{3}]_\psi|$ is maximum ($\sim 20$) around the black contour near the equilateral limit, and it declines as triangles are distorted from this configuration on either direction in $\mu-t$ space. For the obtuse triangles, the $[\alpha_6^{3}]_\xi$ is relatively small, remaining below $1$ when $\xi<0$ (or $\psi>0$) and ranging between $1$ and $4$ when $\xi>0$ (or $\psi<0$). The $|[\alpha_6^{3}]_{b_1}|<2$ for all the cases. The positive $B_6^3$ at equilateral and the negative $B_6^3$ at a few acute triangles below the black contour signal the modification in the gravity.

\begin{figure*}
    \centering 
    \includegraphics[width=.99\textwidth]{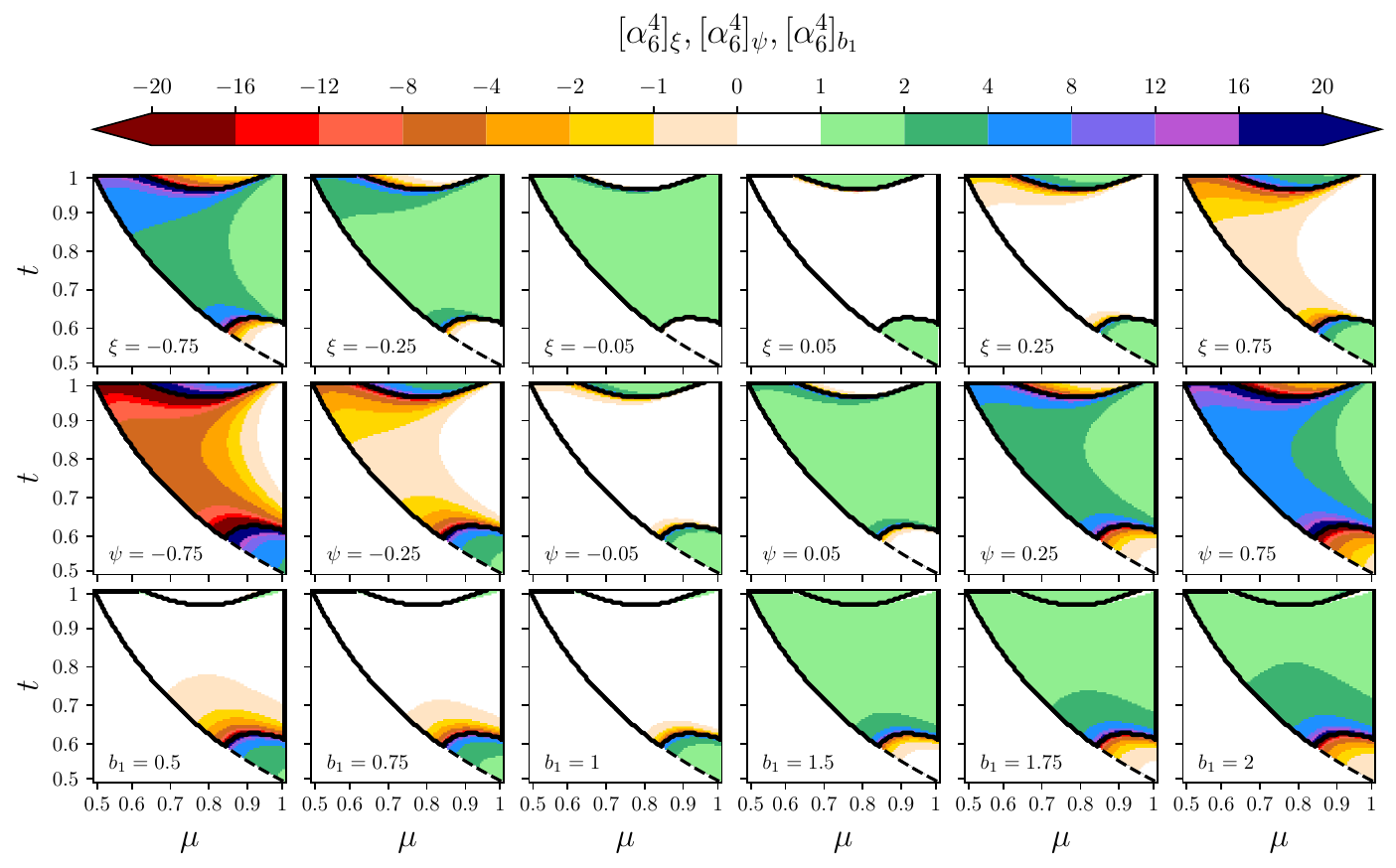} 
    \caption{Sensitivity of $B_6^4$ to variations in the second-order growth indices $\xi$ and $\psi$. The top, middle, and bottom rows correspond to $[\alpha_6^4]_{\xi}$, $[\alpha_6^4]_{\psi}$, and $[\alpha_6^4]_{b_1}$ respectively. The panels are identical to those in Figure~\ref{fig:b60_mut}.}
    \label{fig:b64_mut}
\end{figure*}
Lastly, we discuss the results for $l=6,m=4$ presented in Figure~\ref{fig:b64_mut}. 
Considering the first row, we observe that $[\alpha_6^{4}]_{\xi=-0.75}$ reaches its highest values for equilateral triangles, ranging between $12$ and $16$. It gradually declines as the configuration approaches obtuse triangles but increases again near ($\mu = 0.9$, $t = 0.6$). It remains mostly $2$ to $4$ for the bulk of the triangles near the S-isosceles limit ($2\mu t=1$). Notably, there are a few triangles near the L-isosceles limit ($t=1$) and at ($\mu = 0.9$, $t = 0.6$), where $[\alpha_6^{4}]_\xi \sim -12$. For smaller modification in $\xi$, the magnitude of $[\alpha_6^{4}]_\xi$ decreases significantly. At $\xi = -0.05$, its values remain between $0-2$ and it becomes positive for all triangle configurations when $\xi$ is small (e.g., $\xi = -0.05$ and $\xi = 0.05$). At $\xi = 0.25$, $[\alpha_6^{4}]_\xi$ becomes $\sim -2$ for equilateral triangles, increases to $\sim 8$ for certain L-isosceles triangles, and remains $\sim 2$ near stretched triangle configurations. For the majority of other triangle shapes, $|[\alpha_6^{4}]_\xi|$ remains less than $1$. As $\xi$ approaches $0.75$, the magnitude $|[\alpha_6^{4}]_\xi|$ increases further, reaching a maximum of approximately $\sim 8$ near the equilateral limit, while consistently exhibiting a negative sign for all S-isosceles triangle configurations.  

Considering the second row, we observe that the trend of $[\alpha_6^{4}]_{\psi}$ for $\psi=-0.75$ is similar to that for $\xi = 0.75$, but the maximum value near the equilateral limit exceeds $-20$. The magnitude $|[\alpha_6^{4}]_\psi| \sim 8$ near the stretched limit and increases to $20$ near ($\mu = 0.9$, $t = 0.6$). For the majority of triangle configurations near the S-isosceles limit, $[\alpha_6^{4}]_\psi$ remains mostly between $-4$ and $-8$. the results for $\psi = -0.25$, $-0.05$, $0.05$, and $0.25$ are very similar to those for $\xi = 0.75$, $0.25$, $-0.25$, and $-0.75$, respectively. For $\psi=0.75$, the highest values near equilateral triangles and at ($\mu = 0.9$, $t = 0.6$) range in $12-20$. $[\alpha_6^{4}]_\psi \sim 4-8$ for most of the S-isosceles triangles and $\sim 1-2$ for linear triangles. 

The $b_1$ majorly affects the $B_6^4$ corresponding to stretched triangles and a few obtuse triangles. Considering the third row, we see that the $|[\alpha_6^{4}]_{b_1}|\sim 1-4$ near stretched triangles for $0.5\leq b_1\leq2$. The maximum value of $|[\alpha_6^{4}]_{b_1}|$ is near the black contour around ($\mu = 0.9$, $t = 0.6$), which goes up to $12$ for $b_1=0.5$ and $2$ and decreases to $8$ and $4$ for $b_1=0.75$ and $1$ respectively. For triangles with $t>0.75$, $[\alpha_6^{4}]_{b_1}<1$ when $b_1<1.2$ and $[\alpha_6^{4}]_{b_1}\sim 1-2$ when $b_1>1.2$. The $b_1$ induces the sign flip near stretched triangles and a few obtuse triangles only. Thus, the positive $B_6^4$ values observed for equilateral triangles and for triangles with $0.75 < t < 0.95$, along with the negative $B_6^4$ values for certain L-isosceles triangles (above the black contour), serve as a definitive indicator of modified gravity.

Table~\ref{tab:alpha-sign-flip} summarises the key sign‑flip signatures of bispectrum multipoles $B_l^m$ for testing GR. It encapsulates, for each multipole, the relevant triangle configurations, the contrast between the expected GR sign and the DHOST‑induced flip, and their implications for the second‑order growth indices.

\subsubsection{Higher multipoles}

We find that the bispectrum multipoles above $l=6,m=4$ remain unchanged with respect to $\xi$ and $\psi$. Consequently, $[\alpha_6^5]_\xi=[\alpha_6^6]_\xi=[\alpha_6^5]_\psi=[\alpha_6^6]_\psi=1$ and $[\alpha_8^m]_\xi=[\alpha_8^m]_\psi=1$ (for all $m$) across all triangle configurations. In Equation~\ref{eq:final_bs}, the $\xi$ and $\psi$ dependence is implicit in the kernels $G_{12}$ and $\Delta G_{12}$ which are multiplied by the functions $R$ and $S_{12}$ respectively. The $R$ and $S_{12}$, which are polynomials of the order $\mu^6_i$ and $\mu^4_i$ respectively, only contribute to the multipoles $l\leq 6$ and $l\leq 4$ respectively. As a result, the $\xi$ dependence is not captured by $l>6$ multipoles.

\begin{table*}[ht]
  \centering
  \caption{ Summary of the sign-flip features observed in bispectrum multipoles $B_l^m$ as indicators of deviations from GR. The first column lists the multipoles exhibiting prominent sign-flip behaviour. The second column specifies the triangle configurations associated with the sign flip for each multipole. The third column details the direction of the sign flip, contrasting the expected sign under GR with the altered sign induced in the DHOST scenario.  The fourth column presents the physical interpretation of the sign flip and its implications for the second-order growth indices.}
  \label{tab:alpha-sign-flip}
  \begin{tabular}{l|l|l|l}
    \toprule
    Multipole $B_l^m$                         & Shape   ($\mu,t$)           & Sign flip direction       & Indication \\
                            &         & (GR\,$\to$\,DHOST) &                    \\
    \midrule
    $B_2^1$                          & acute \& equilateral    & \hspace{.23cm} $+\to -$        & $\xi<\xi_{\rm GR}$ or $\psi>\psi_{\rm GR}$             \\
                                     & close to right angle& \hspace{.23cm} $-\to +$        & $\xi>\xi_{\rm GR}$ or $\psi<\psi_{\rm GR}$             \\
    \Gmidrule
    $B_4^3$                          & acute \& stretched& \hspace{.23cm} $+\to -$        & $\xi<\xi_{\rm GR}$ or $\psi>\psi_{\rm GR}$             \\
                                     & obtuse except stretched limit& \hspace{.23cm} $-\to +$        & $\xi>\xi_{\rm GR}$ or $\psi<\psi_{\rm GR}$             \\
    \Gmidrule
    $B_6^0$                          &  equilateral  & \hspace{.23cm} $-\to +$         & $\xi>\xi_{\rm GR}$ or $\psi<\psi_{\rm GR}$            \\
                                     & acute except equilateral limit    & \hspace{.23cm} $+\to -$       & $\xi<\xi_{\rm GR}$ or $\psi>\psi_{\rm GR}$            \\
    \Gmidrule
    $B_6^1$                          & all acute triangles & \hspace{.23cm} $-\to +$      & $\xi<\xi_{\rm GR}$ or $\psi>\psi_{\rm GR}$             \\
    \Gmidrule
    $B_6^2$                          & equilateral       & \hspace{.23cm} $+\to -$        & $\xi>\xi_{\rm GR}$ or $\psi<\psi_{\rm GR}$             \\
                                     & acute except equilateral limit & \hspace{.23cm} $-\to +$    & $\xi<\xi_{\rm GR}$ or $\psi>\psi_{\rm GR}$             \\
    \Gmidrule
    $B_6^3$                          & equilateral \& squeezed limit      & \hspace{.23cm} $-\to +$        & $\xi>\xi_{\rm GR}$ or $\psi<\psi_{\rm GR}$             \\
                                     & acute except equilateral limit & \hspace{.23cm} $+\to -$    & $\xi<\xi_{\rm GR}$ or $\psi>\psi_{\rm GR}$             \\
    \Gmidrule
    $B_6^4$                          & stretched \& L‑isosceles   & \hspace{.23cm} $+\to -$& $\xi<\xi_{\rm GR}$ or $\psi>\psi_{\rm GR}$             \\
                                     & equilateral \& S‑isosceles  & \hspace{.23cm} $-\to +$        & $\xi>\xi_{\rm GR}$ or $\psi<\psi_{\rm GR}$             \\
    \bottomrule
  \end{tabular}
\end{table*}




\section{Summary and discussion}
\label{sec:summary}

In this work, we have explored the potential of the redshift space bispectrum multipole moments of biased tracers as a tool for probing modified gravity theories beyond linear approximation. Here we employ the standard perturbation theory, incorporating second-order perturbations in both the density and velocity fields to capture higher-order properties of gravitational interactions. The analysis is conducted within the framework of the most general scalar-tensor theory of gravity, known as the DHOST theory \citep{Langlois:2018dxi,Kobayashi:2019hrl}. The influence of modified gravity under the DHOST framework is characterised by three key parameters: a linear growth index, $\gamma$, and two second-order growth indices, $\xi$ and $\psi$, which encapsulate additional information about structure growth beyond the scope of linear theory \citep{Yamauchi:2017ibz,Yamauchi:2021arXiv210802382Y,Sugiyama:2023MNRAS.523.3133S}. For the standard GR, these parameters take the values $\gamma_\text{GR} = 6/11$, $\xi_\text{GR} = 3/572$, and $\psi_\text{GR} = 0$. Any deviation from these values indicates a departure from GR. 

Modifications to gravity affect the clustering properties of biased tracers and the coupling between modes. The bispectrum, being sensitive to the non-linearity and mode coupling introduced by gravitational evolution, serves as a powerful probe for detecting deviations from GR \citep{Peebles:1980lssu.book.....P,Scoccimarro:2000ApJ...544..597S,Sugiyama:2023MNRAS.523.3133S}. These signatures of modified gravity in the bispectrum become even more pronounced in redshift space, where the anisotropies introduced by peculiar velocities are sensitive solely to the theory of gravity, independent of tracer bias. We decompose the anisotropic redshift space bispectrum in the spherical harmonic basis $Y_l^m(\pp)$ \citep{Scoccimarro:1999ed,Bharadwaj:2020MNRAS.493..594B} and study the sensitivity of all possible bispectrum multipole moments $B_l^m(k_1,\mu,t)$ to modifications in gravity across all triangle configurations ($\mu,t$). 

We show that the higher-order multipoles, with $l=2,4,6$, are comparatively more sensitive to modified gravity than the spherically averaged $l=0$ moment, which has predominantly been the focus of prior studies. Considering the linear growth, the $B_0^0$ remains close to the values predicted by the GR for a range of $\gamma$. This is consistent with the earlier findings \citep{Borisov:2009PhRvD..79j3506B,Bernardeau:2011JCAP...06..019B,Tatekawa:2008JCAP...09..009T}. The percent level measurements are required to detect deviations in $\gamma$ using the monopole. Similarly, the $l=2$ multipoles also show limited sensitivity to the $\gamma$. In contrast, the $l=6$ multipoles, particularly for equilateral and squeezed triangles, are more sensitive to the $\gamma$. In addition, the $B_4^2$ and $B_4^3$ multipoles for stretched triangles show high sensitivity to the $\gamma$. Notably, the highest non-zero multipole is $(l=8, m=6)$. However, all $(l=8)$ multipoles do not provide any independent information on $\gamma$ beyond what is already captured by the multipoles with $l < 8$.

The higher-order multipoles exhibit exceptional sensitivity to the non-linear growth indices, $\xi$ and $\psi$. Specifically, the multipoles $B_2^1$, $B_4^3$, $B_6^0$, $B_6^1$, $B_6^2$, $B_6^3$, and $B_6^4$ encapsulate substantial information about departures from GR. The sensitivity of the multipoles varies with triangle configurations. The acute ($t > \mu$) triangle configurations, in general, demonstrate the strongest sensitivity to $\xi$ and $\psi$. For $B_2^1$ and $B_4^3$, the sensitivity peaks near right-angle triangle configurations. In contrast, for the $l=6$ multipoles, equilateral triangles are the most sensitive, with sensitivity generally decreasing as triangles become more deformed towards squeezed or stretched shapes. An exception is $B_6^4$, which shows high sensitivity for stretched triangles as well. We find that the $(l=6,m>4)$ and all $l=8$ multipoles are unable to capture second-order growth.

We demonstrate that, for specific triangular configurations, the values of these multipoles exhibit opposite signs in modified gravity compared to those predicted by GR. This sign reversal serves as a robust indicator of deviations from GR. The characteristic signatures of modified gravity are as follows: for $B_2^1$, negative values for acute triangles and positive values near right-angle triangles; for $B_4^3$, negative values for acute and stretched triangles and positive values for obtuse triangles; for $B_6^0$, negative values for acute triangles and positive values for equilateral triangles; for $B_6^1$, positive values for all acute triangles; for $B_6^2$, negative values for equilateral triangles and positive values for other acute triangles; for $B_6^3$, positive values near the equilateral limit and negative values for other acute triangles; and for $B_6^4$, negative values for many L isosceles and stretched triangles and positive values near the S isosceles limit. These distinct signatures, tabulated in Table \ref{tab:alpha-sign-flip}, underscore the utility of higher-order multipoles in probing deviations from GR.

Although various bispectrum multipoles effectively capture robust signatures of modified gravity, the linear bias $b_1$ and quadratic bias parameters $b_2$ and $b_{s^2}$ have the tendency to replicate these signals if they are not correctly accounted for. While the impact of bias on $B_2^1$ is minimal, it is significantly more pronounced for the case of $B_4^3$. We show that the $l=6$ multipoles offer a distinct advantage over $l=2$ and $l=4$ multipoles as they are not affected by the quadratic bias $b_2$ and second-order tidal bias $b_{s^2}$. Consequently, $l=6$ multipoles hold considerable potential for probing and constraining theories of modified gravity, emphasising the need to leverage their capabilities in analyses. However, it remains crucial to recognise that similar bispectrum features may arise from either bias or modified gravity. Therefore, a careful disentanglement of these contributions is essential to reliably constrain both the gravitational theory and bias parameters of tracers in the analysis. The ongoing and future surveys such as Euclid\footnote{https://www.euclid-ec.org/} \citep{Laureijs_2011}, DESI\footnote{https://www.desi.lbl.gov/}\citep{Levi_2013}, and SKA\footnote{http://www.skatelescope.org}\citep{Weltman:2020PASA...37....2W} will provide us with an unprecedented volume of data, enabling the detection of modified gravity signatures and constraining gravitational theories with percent-level precision. In future work, we will utilise the fast estimator \citep{Gill:2024arXiv240514513S} to compute these higher-order bispectrum moments from galaxy survey data and extract critical information about the underlying theory of gravity.

\section*{Acknowledgements}

The author is grateful to Barbie Sangha for helpful discussions and encouragement throughout this project. We thank the anonymous reviewers for their valuable comments.

\section*{Data availability}
The codes and packages for numerical computation involved in this work will be shared on reasonable request to the author.

\bibliography{main}

\appendix
\section{Results for $l=0$ and other $l=2,4$ multipoles}
\label{sec:appendix}
\begin{figure*}
    \centering 
    \includegraphics[width=.99\textwidth]{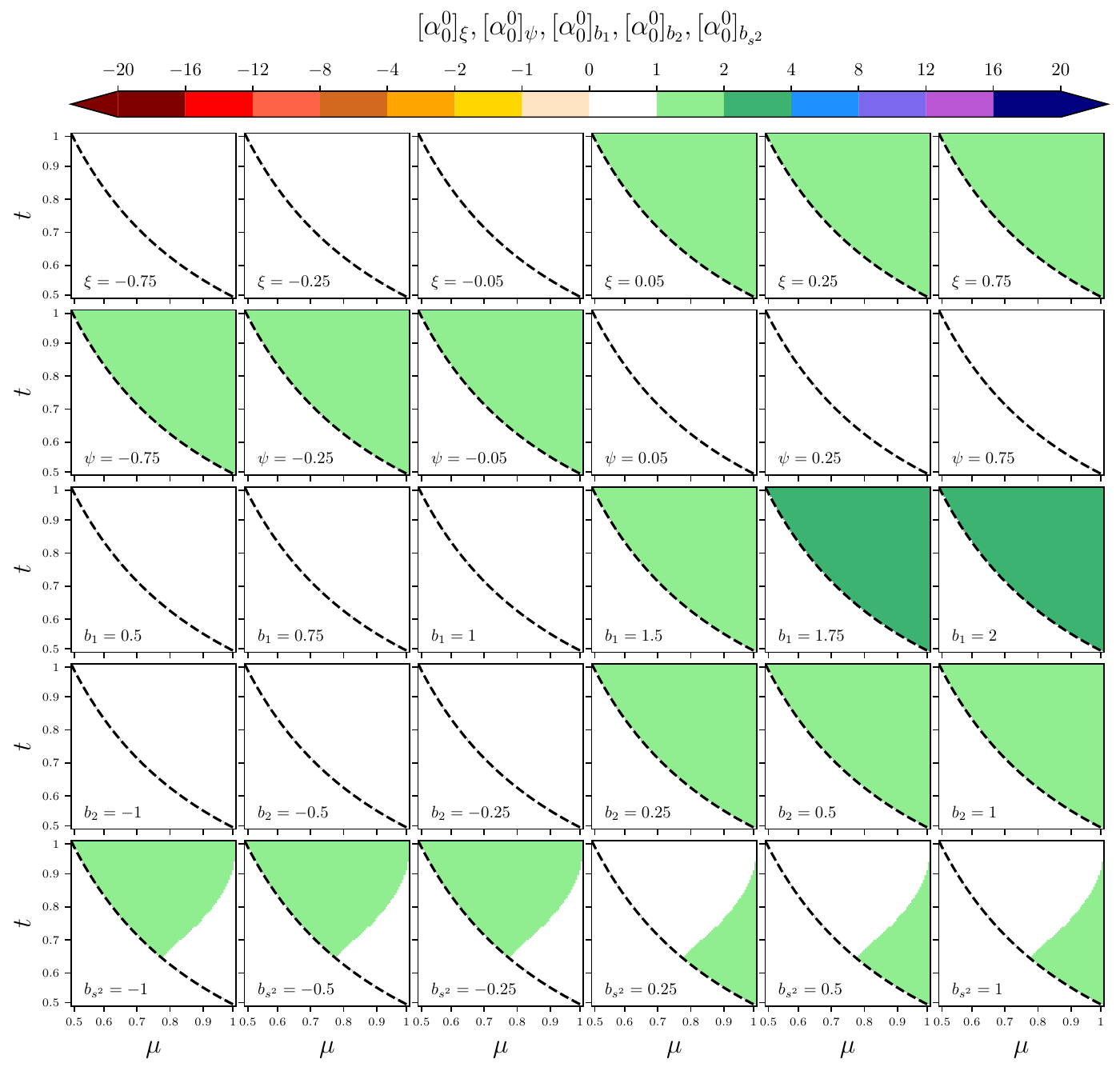} 
    \caption{Sensitivity of $B_0^0$ to variations in the second-order growth indices $\xi$ and $\psi$. The panels are identical to those in  Figure~\ref{fig:b21_mut}.}
    \label{fig:b00}
\end{figure*}

\begin{figure*}
    \centering 
    \includegraphics[width=.99\textwidth]{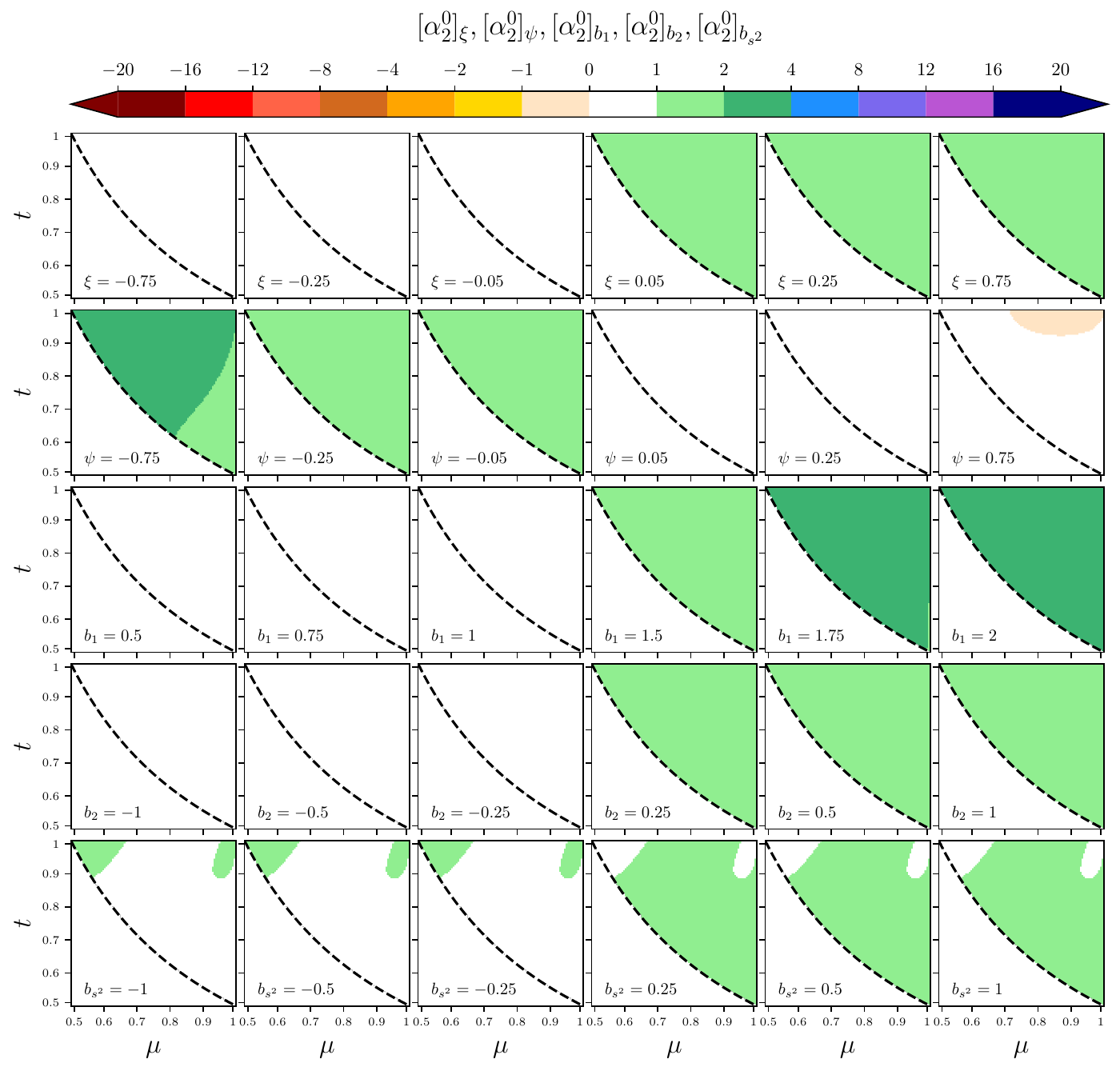} 
    \caption{Sensitivity of $B_2^0$ to variations in the second-order growth indices $\xi$ and $\psi$. The panels are identical to those in  Figure~\ref{fig:b21_mut}.}
    \label{fig:b20}
\end{figure*}

\begin{figure*}
    \centering 
    \includegraphics[width=.99\textwidth]{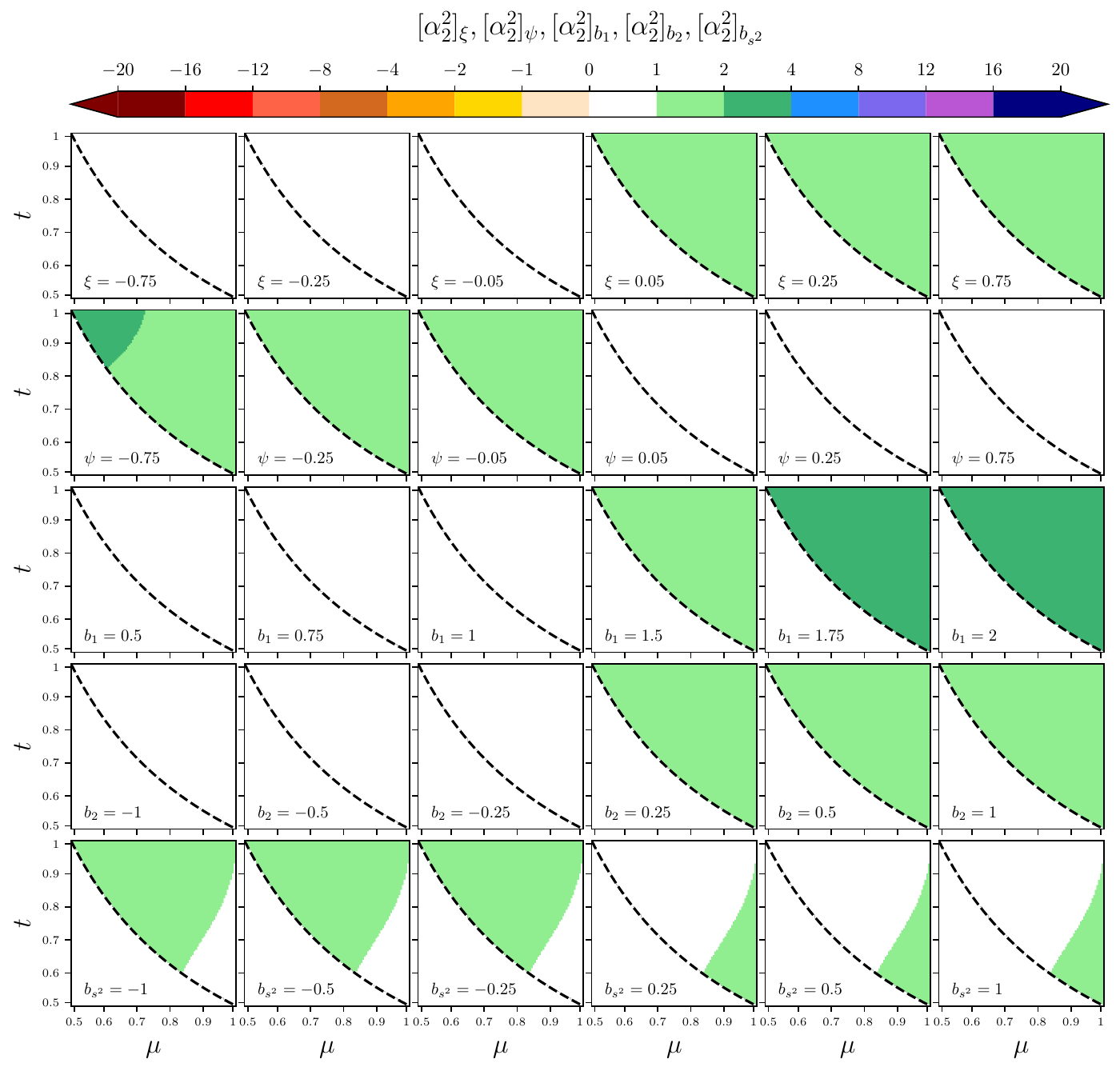} 
    \caption{Sensitivity of $B_2^2$ to variations in the second-order growth indices $\xi$ and $\psi$. The panels are identical to those in  Figure~\ref{fig:b21_mut}.}
    \label{fig:b22}
\end{figure*}

\begin{figure*}
    \centering 
    \includegraphics[width=.99\textwidth]{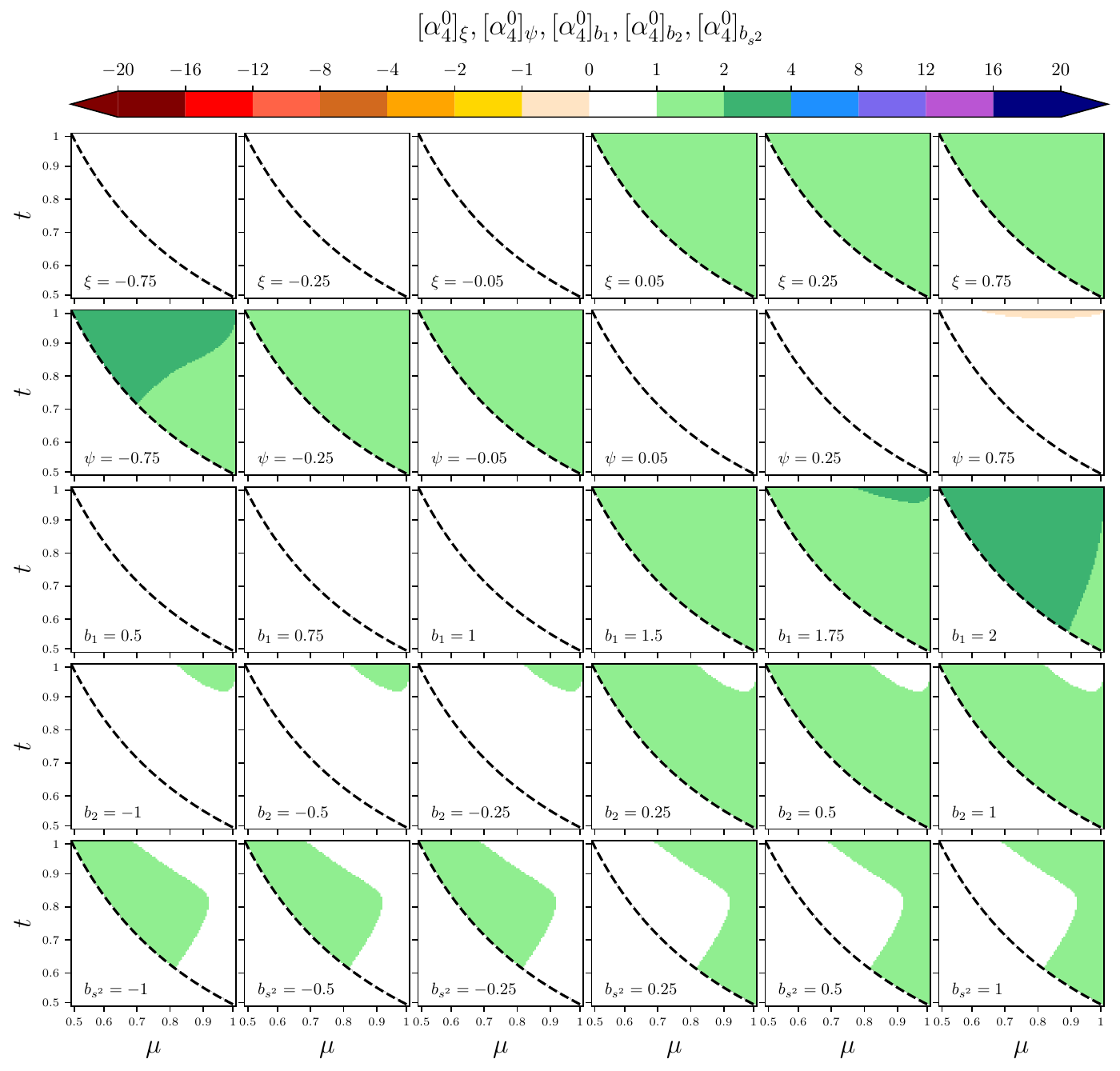} 
    \caption{Sensitivity of $B_4^0$ to variations in the second-order growth indices $\xi$ and $\psi$. The panels are identical to those in  Figure~\ref{fig:b21_mut}.}
    \label{fig:b40}
\end{figure*}

\begin{figure*}
    \centering 
    \includegraphics[width=.99\textwidth]{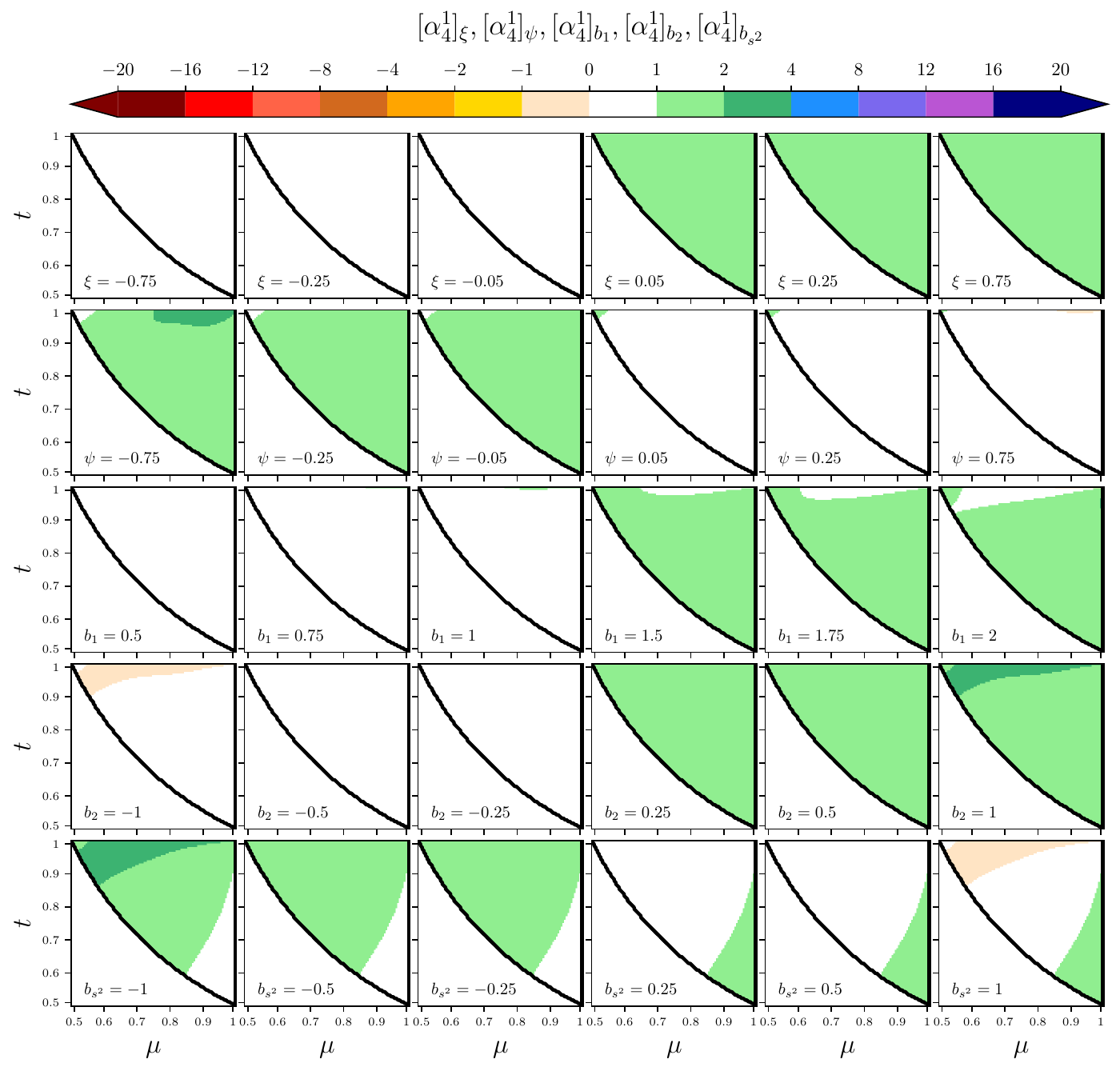} 
    \caption{Sensitivity of $B_4^1$ to variations in the second-order growth indices $\xi$ and $\psi$. The panels are identical to those in  Figure~\ref{fig:b21_mut}.}
    \label{fig:b41}
\end{figure*}

\begin{figure*}
    \centering 
    \includegraphics[width=.99\textwidth]{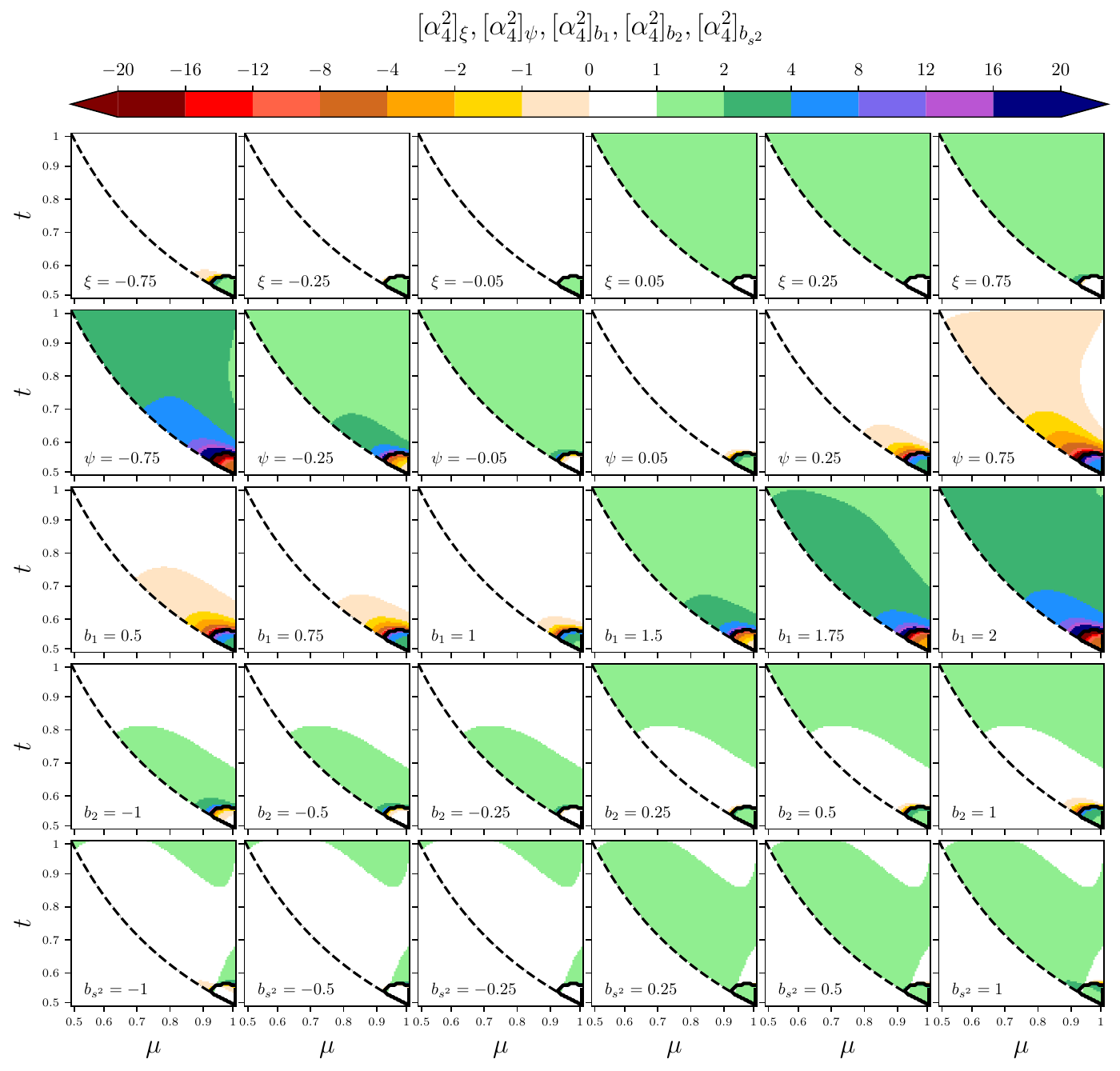} 
    \caption{Sensitivity of $B_4^2$ to variations in the second-order growth indices $\xi$ and $\psi$. The panels are identical to those in  Figure~\ref{fig:b21_mut}.}
    \label{fig:b42}
\end{figure*}

\begin{figure*}
    \centering 
    \includegraphics[width=.99\textwidth]{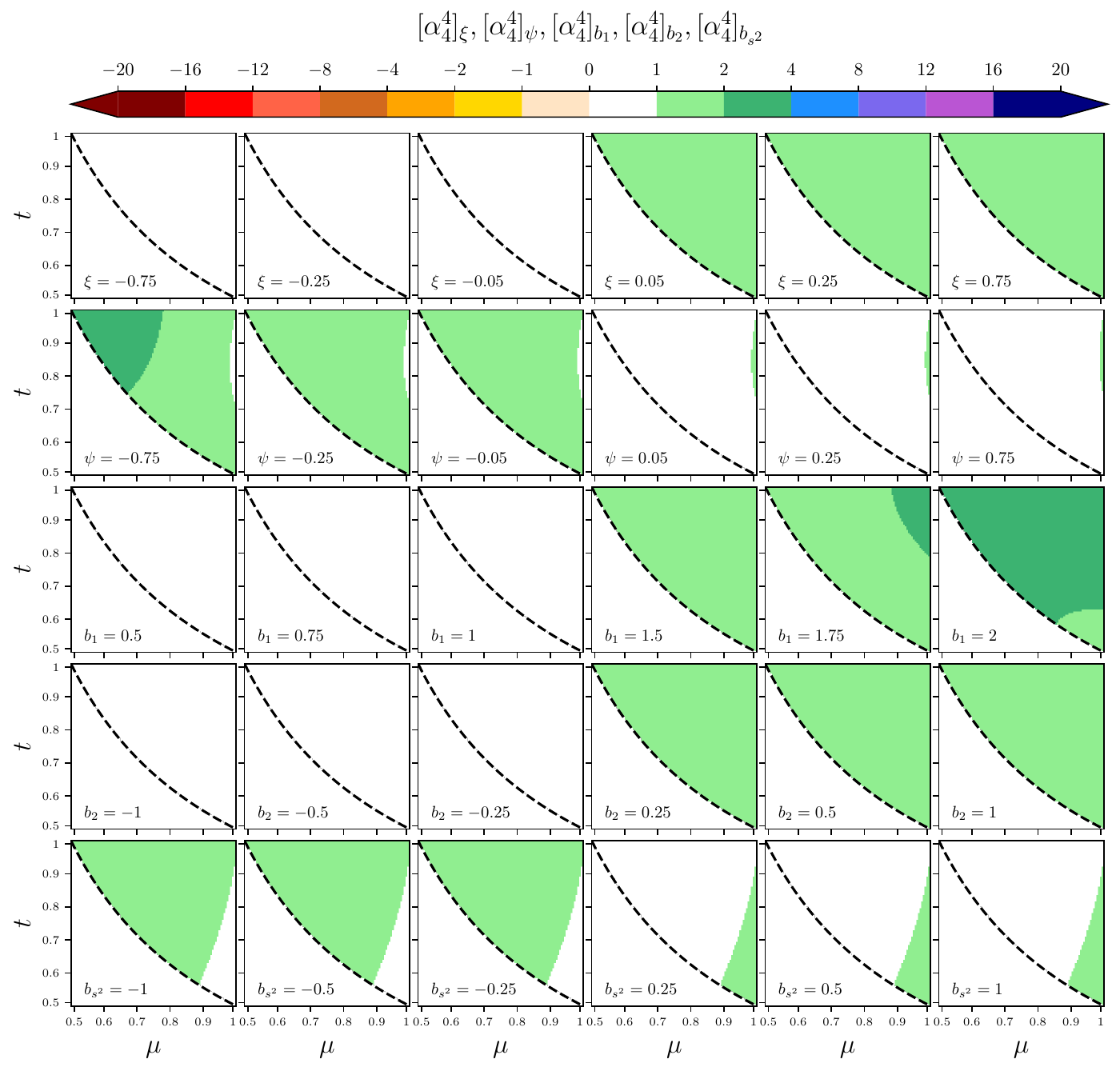} 
    \caption{Sensitivity of $B_4^4$ to variations in the second-order growth indices $\xi$ and $\psi$. The panels are identical to those in  Figure~\ref{fig:b21_mut}.}
    \label{fig:b44}
\end{figure*}

Figures ~\ref{fig:b00}, \ref{fig:b20}, \ref{fig:b22}, \ref{fig:b40}, \ref{fig:b41}, \ref{fig:b42}, and \ref{fig:b44} shows the sensitivity of $B_0^0$, $B_2^0$, $B_2^2$, $B_4^0$, $B_4^1$, $B_4^2$, and $B_4^4$, respectively, to the variations in second-order growth indices $\xi$ and $\psi$. The results indicate that these multipoles exhibit minimal variation with changes in gravity at the second order. The ratio $[\alpha_l^m]_\xi$ and $[\alpha_l^m]_\psi$ remain below 2 in most cases, although they can reach values of up to $\sim 4$ for a few acute triangle configurations, especially for $B_2^0,B_2^2,B_4^0,B_4^2,B_4^4$. Notably, there is no consistent indication of a sign flip signalling the modified gravity in these multipoles, except for $B_4^2$ which show sign change near triangles near $\mu\approx 0.9, t\approx=0.6$ for $\psi>0$ (or $\xi<0$) and near stretched triangles for $\psi<0$ (or $\xi>0$). Additionally, the signal of modified gravity in these multipoles is strongly influenced by the bias parameters $b_1$, $b_2$, and $b_{s^2}$.

\end{document}